\documentclass[a4paper,fleqn,usenatbib]{mnras}

\usepackage[T1]{fontenc}
\usepackage{ae,aecompl}

\usepackage{graphicx}	
\usepackage{amsmath}	
\usepackage{amssymb}	

 \title[IGR J17091--3624 - \textit{AstroSat} view]{ \textit{AstroSat} view of IGR J17091--3624 and 
GRS 1915+105: decoding the `pulse' in the `Heartbeat State'}

\author[Tilak Katoch et al.]
{Tilak Katoch$^{1}$\thanks{E-mail: tilak@tifr.res.in},
Blessy E. Baby$^{2,3}$,
Anuj Nandi$^{2}$,
V. K. Agrawal$^{2}$,
H. M. Antia$^{1}$ and
\newauthor
Kallol Mukerjee$^{1}$.
\\
$^{1}$Department of Astronomy \& Astrophysics, Tata Institute of Fundamental Research, Homi Bhabha Road, Colaba, Mumbai, 400005, India\\
$^{2}$Space Astronomy Group, ISITE Campus, U. R. Rao Satellite Centre, Outer Ring Road, Marathahalli, Bangalore, 560037, India\\
$^{3}$Department of Physics, University of Calicut, Malappuram, Kerala, 673635, India.
}

\date{Accepted XXX. Received YYY; in original form ZZZ}

\pubyear{2020}

\begin{document}
\label{firstpage}
\pagerange{\pageref{firstpage}--\pageref{lastpage}}
\maketitle

\begin{abstract}
IGR J17091--3624 is a transient galactic black hole which has a distinct quasi-periodic variability known as `heartbeat', similar to the one observed in GRS 1915+105. In this paper, we report the results of $\sim 125$ ks \textit{AstroSat} observations of this source during the 2016 outburst. For the first time a double peaked QPO (DPQ) is detected in a few time segments of this source with a difference of $\delta f ~\sim12$ mHz between the two peaks. The nature of the DPQ was studied based on hardness ratios and using the static as well as the dynamic power spectrum. Additionally, a low frequency (25--48 mHz) `heartbeat' single peak QPO (SPQ) was observed at different intervals of time along with harmonics ($50-95$ mHz). Broadband spectra in the range $0.7-23$ keV, obtained with \textit{SXT} and \textit{LAXPC}, could be fitted well with combination of a thermal Comptonisation and a multicolour disc component model. During \textit{AstroSat} observation, the source was in the Soft-Intermediate State (SIMS) as observed with \textit{Swift/XRT}. We present a comparative study of the `heartbeat' state variability in IGR J17091--3624 with GRS 1915+105. Significant difference in the timing properties is observed although spectral parameters ($\Gamma\sim2.1-2.4$ and $T_\mathrm{max}\sim0.6-0.8$ keV) in the broad energy band remain similar. Spectral properties of segments exhibiting SPQ and DPQ are further studied using simple phase resolved spectroscopy which does not show a significant difference. Based on the model parameters, we obtain the maximum ratio of mass accretion rate in GRS 1915+105 to that in IGR J17091--3624 as $\sim25:1$. We discuss the implications of our findings and comment on the physical origin of these exotic variabilities.

\end{abstract}



\section{Introduction}

A Low Mass X-ray Binary (LMXB) is formed when matter is accreted from a nearby companion star onto a compact object such as a black hole (BH) or a neutron star (NS) \citep{Frank2002}. They are categorized as transient and persistent sources based on their outbursts studied over a period of time. Most of the LMXBs display sudden outbursts marked by increase in luminosity by a few orders of magnitude, within tens of days, and remain active for several weeks or months before decaying to quiescent phase \citep{McC2010}. It has been noticed from previous studies that the timing and spectral properties of some of the Black Hole Binaries (BHBs) display similar features during the outbursts \citep[and references therein]{Miya1992, Psalt1999, Mend2001, Bell2011, Ingram2020}. 

The accretion dynamics close to the black hole can be studied by investigating the temporal and spectral properties of BHBs during the outbursts. The outburst begins in the hard state and thereafter transits to intermediate and high-soft state (HSS) resulting in softening of the spectrum. The source then traverses back to the hard state through intermediate states before decaying to quiescent phase \citep{Bell2005,Nandi2012,Nandi2018,Sreehari2019}. Strong broadband noise along with Quasi-periodic Oscillations (QPOs) are seen in the Power Density Spectra (PDS) during the low-hard state (LHS) and hard-intermediate state (HIMS) while the PDS of soft-intermediate state (SIMS) and HSS are characterized by a red noise and weak/no signature of QPO features. The energy spectra are usually characterized by variations in the relative contributions of an optically thick disc and an optically thin corona produced by Comptonisation of seed photons by hotter electrons \citep{Tana1995, Chak1995}.

IGR J17091--3624 is known to be a galactic black hole (GBH) in a LMXB system. It was first discovered by \textit{INTEGRAL/IBIS} in April 2003 \citep{kuul2003}. The source was in outburst in 1994, 2001, 2003, 2007, 2011 and 2016 \citep[e.g.,][]{Capt2009, Krim2011, Mill2016, Sree2018} with
a quiescent period of typically four years between these outbursts. It is observed that these bursts last for a couple of months. Although, mass, inclination and distance are not known precisely due to the non-detection of an optical counterpart, approximate estimates of these parameters exist through other means. Distance is estimated to be in the range of $ 11-17 $ kpc \citep{Rodri2011} using the luminosity at hard to soft state transition assuming mass of the black hole to be $10~M_{\sun}$ \citep[see][]{Iyer2015}. Better constraints on mass were obtained by \cite{Radhi2018} using two component flow model as $10.6-12.3$~$M_{\sun}$. \citet{King2012} have proposed an inclination angle between $50^{\circ}$ to $70^{\circ}$, the upper limit being determined by the absence of any eclipses. 

The most striking feature of the source is the detection of variability classes similar to those observed in GRS 1915+105, discovered a decade earlier by \textit{WATCH} onboard \textit{GRANAT} \citep{Castro1992}. GRS 1915+105 contains one of the most massive black holes among all the stellar mass black hole systems discovered so far with an estimated mass of $12.4^{+2.4}_{-1.8} M_{\odot}$ lying at a distance of $\sim 8.5^{+2.0}_{-1.6}$ kpc \citep{Reid2014}. The light curve of this source shows a very rich class of variability pattern which have been categorized into 14 different classes based on its light curve and colour-colour diagram \citep{Belloni2000, Klein-Wolt2002, Hannikainen2005}. The exhibition of variability on time-scales from seconds to minutes makes it an interesting case study. Of all the variability classes, the $\rho$-class or `heartbeat' State (HS) is the most widely studied due to its uniquely structured, regular periodic profile in the X-ray light curve \citep{Neilsen2011, Mine2012}. Comparative studies for the two similar classes of these two different sources --- IGR J17091--3624 and GRS 1915+105, provide an excellent opportunity to decipher the underlying physics behind them.

IGR J17091--3624 shows morphological similarities with GRS 1915+105. During the outburst, IGR J17091--3624 revealed 9 set of variability classes of which seven are analogous to different classes observed in GRS 1915+105 \citep{Court2017} and two are unique to the source. 
The most prominent pattern of the IGR J17091--3624 was the `heartbeat' oscillations, similar to the $\rho$-class in GRS 1915+105. A high-frequency QPO (HFQPO) having a frequency of 66 Hz was reported by \cite{Alta2012}, identical to the one observed in GRS 1915+105 at $ \sim67 $ Hz \citep{Morg1997}. A low frequency feature was also noticed during the 2011 outburst as reported by \cite{Alta2011} with a frequency of 10 mHz initially which then evolved to QPOs in the $ 25-30 $ mHz range in the later days. A similar low frequency feature was reported from $\sim 20$ s to 70 s in `heartbeat' state \citep{Capi2012}. \cite{Pahari2014} also discovered an intermediate class before the `heartbeat' class which showed variabilities between $ 11-14 $ mHz and a broad range of low frequency features between $\sim 10$ to 100 mHz including the harmonics during the $\rho$-class \citep{Zhang2014}. In the next outburst of 2016, such low frequency features were reported by \cite{Radhi2018} in the range of $ 20-30 $ mHz. These low frequency features are considered similar to those seen in GRS 1915+105 resulting from the structured variability.

In this paper, we present the analysis of 2016 outburst of the IGR J17091--3624 observations by \textit{AstroSat} using data from \textit{SXT} and \textit{LAXPC} instruments. A study of variabilities was carried out in the form of static and dynamic Power Density Spectrum (PDS) to look at low frequency QPOs during different phases of the bursts. We did not find any transition of states during the entire observation as the source was in soft-intermediate state \citep{Radhi2018}. We found the aperiodic burst ($ 25-40 $ mHz) resembling `heartbeat' oscillation with harmonics varying between $ 55-95 $ mHz, which are referred to as the `heartbeat' QPOs \citep{Alta2011, Pasham2013}. For the first time, we observe a distinct split in this feature varying at two different frequency bands $ 34.0-36.8 $ mHz and $ 44.1-48.2 $ mHz in the static as well as dynamic PDS of a few segments. A very low ($ 5-10 $ mHz) frequency feature was also found along with its harmonics. Finally, the broadband spectral analysis of the source was carried out over the limited energy range $0.5-23$ keV using the \textit{SXT} and \textit{LAXPC} data due to the presence of another source, GX 349+2, in the \textit{LAXPC} FOV (see, Sect.~\ref{sec:cont} for details). \textit{LAXPC} data in $3-20$ keV was used for temporal analysis to lend more credibility to our findings with \textit{SXT} data owing to its better temporal resolution. We also perform a comparative study of IGR J17091--3624 with respect to the temporal and spectral results obtained from the \textit{AstroSat} observations of GRS 1915+105 in a similar state.

A summary of the procedures followed for data reduction is given in Sect.~\ref{obs}. The methodology considered for analysis of data from \textit{SXT} and \textit{LAXPC} are discussed in Sect.~\ref{ana}. The results obtained from the temporal and spectral analysis using phenomenological models and comparison with results from analysis of GRS 1915+105 are presented in Sect.~\ref{res}. Finally, we discuss the implications of our results and summarize our findings in  Sect.~\ref{dis}.

\section{Observations \& Data Reduction}
 \label{obs}

The Indian multi-wavelength observatory, \textit{AstroSat}, has the capability to view the sky in the X-ray band from 0.3 to 100 keV simultaneously. This can be achieved with the help of on-board co-aligned X-ray instruments --- \textit{Soft X-ray Telescope (SXT)} \citep{singh2017}, \textit{Large Area X-ray Proportional Counter (LAXPC)} \citep{Antia2017} and \textit{Cadmium Zinc Telluride Imager (CZTI)} \citep{vada2016}. \textit{AstroSat} observed IGR J17091--3624 between Apr 26, 2016 (MJD 57504.26) and Apr 27, 2016 (MJD 57505.72) for the duration of $\sim125$ ks during a ToO observation\footnote{ObsID: 20160426\_T01\_118T01\_9000000430}. We found 35 segments of Good Time Intervals (GTIs) out of which 19 segments $\ge$ 1 ks were considered having the total duration of $\sim 37$ ks, while the rest of GTIs were not considered due to an exposure time $\le1$ ks or presence of data gaps. The \textit{AstroSat} data from the \textit{SXT} and \textit{LAXPC} instruments were used for the timing and spectral analysis. During this observation, these instruments were in their default operational modes of Photon Counting (PC) mode (\textit{SXT}) and Event Analysis (EA) mode (\textit{LAXPC}). \textit{SXT} Level-2 data were obtained from the ISSDC data dissemination archive\footnote{\url{http://astrobrowse.issdc.gov.in/astro\_archive/archive}}. The entire \textit{SXT} observation of IGR J17091--3624 is summarized in the Table~\ref{tab:1} where all the 19 segments with their exposure times and average count rates are also listed.

\begin{table}
\centering
\caption{The table describes two \textit{AstroSat} observations. The top section describes the IGR J17091--3624 observation (ObsID: 20160426\_T01\_118T01\_9000000430) details of the \textit{SXT} ($0.5-7$ keV) segments used for this analysis. The bottom section gives details of GRS 1915+105 observation (ObsId: 20170328\_G06\_033T01\_9000001116).}
 \label{tab:1}
	\begin{tabular}{ccc} 
		\hline
		\multicolumn{3}{c}{IGR J17091--3624$^{*}$}\\
		\hline
		Segment & Exposure & SXT Avg. Rate\\
		(No)    &  (s)     & (counts s$^{-1}$)\\
		\hline	 
		01 & 1749 & $12.79 \pm 0.09$\\
		02 & 2275 & $12.60 \pm 0.07$\\
		03$\dagger$ & 2289 & $12.67 \pm 0.07$\\
		04 & 2291 & $12.78 \pm 0.08$\\
		05 & 2291 & $12.53 \pm 0.07$\\
		06 & 1483 & $12.24 \pm 0.09$\\
		07$\dagger$ & 2292 & $12.44 \pm 0.07$\\ 
		08 & 2294 & $12.53 \pm 0.07$\\
		09 & 1507 & $12.25 \pm 0.09$\\
		10 & 1671 & $12.10 \pm 0.09$\\
		11$\dagger$ & 1244 & $11.74 \pm 0.10$\\
		12 &  993 & $12.28 \pm 0.11$\\
		13 & 1324 & $12.19 \pm 0.10$\\
		14 & 1455 & $12.42 \pm 0.09$\\
		15 & 2294 & $12.47 \pm 0.07$\\
		16$\dagger$ & 2296 & $12.40 \pm 0.07$\\
		17 & 2296 & $12.42 \pm 0.07$\\
		18 & 2296 & $12.51 \pm 0.07$\\
		19 & 2296 & $12.24 \pm 0.07$\\	
		\hline
		\multicolumn{3}{c}{GRS 1915+105}\\
		\hline
		Orbit & Exposure & Avg. Rate\\
		(No)  &  (s)     & (counts s$^{-1}$)\\
		\hline
		8118 & 1201 & $23.75 \pm 0.14$ (SXT)\\
		     & 1774 & $771.50 \pm 0.67$ (LAXPC)\\
		\hline
		\multicolumn{3}{c}{\textsuperscript{*}\footnotesize{Observation log of \textit{LAXPC} data of IGR J17091--3624 is not}}\\
		\multicolumn{3}{c}{\footnotesize{mentioned due to contamination by the source GX 349+2.}}\\
		\multicolumn{3}{c}{\footnotesize{See section \ref{sec:cont} for details. Rows marked with $\dagger$ are the}} \\
		\multicolumn{3}{c}{\footnotesize{ones selected for the broadband spectral analysis.}}
    \end{tabular}
\end{table}

\textit{AstroSat} observed the source GRS 1915+105 for a total duration of $\sim 90$ ks on 28 March 2017 as part of a Guaranteed Time (GT) proposal\footnote{ObsId: 20170328\_G06\_033T01\_9000001116}. The source was found to exhibit three different variability classes in this time period \citep{Rawat2019}. It was in the $\chi$ class at the beginning, before undergoing a transition to the `heartbeat' state through an intermediate state. `Heartbeat' state was observed continuously for four orbits during this observational campaign, which amounts to an effective exposure time of 10 ks. Of these 4 orbits, we choose the orbit with longest uninterrupted GTI for both instruments. Archival data of both \textit{SXT} and \textit{LAXPC} have been used for broadband timing and spectral analysis. The net exposure time for the instruments were 1201 s and 1774 s, respectively (see Table \ref{tab:1}). \textit{SXT} was operated in the Photon Counting (PC) Mode for this duration while all three \textit{LAXPC} units were in the Event Analysis Mode (EA).

\subsection{SXT Data Reduction}

For the \textit{SXT} data analysis, we follow the guidelines given in the \textit{SXT} user manual available at the website\footnote{\url{http://www.tifr.res.in/\~astrosat\_sxt/index.html}}. We incorporated the Level-2 data generated from the \textit{SXT} pipeline version 1.4b released on 03 January 2019. The data analysis tools are downloaded from the \textit{SXT} payload operation centre website along with background, spectral response and auxiliary response files (ARF) as suggested by the \textit{SXT} instrument team for the analysis. All individual orbit data from PC mode were merged to generate the single clean event file using the \textit{SXT} event merger script (\texttt{Julia v1.0}). Thereafter, the cleaned event file was used to generate the light curve and spectra between $ 0.5-7 $ keV using \texttt{XSELECT 2.4g}. Offset ARF was generated for this data, to correct for the offset between \textit{SXT} and \textit{LAXPC} instruments, as the source pointing of the satellite was made with respect to \textit{LAXPC} instrument. 

\begin{figure*}
	{\includegraphics[scale=0.39]{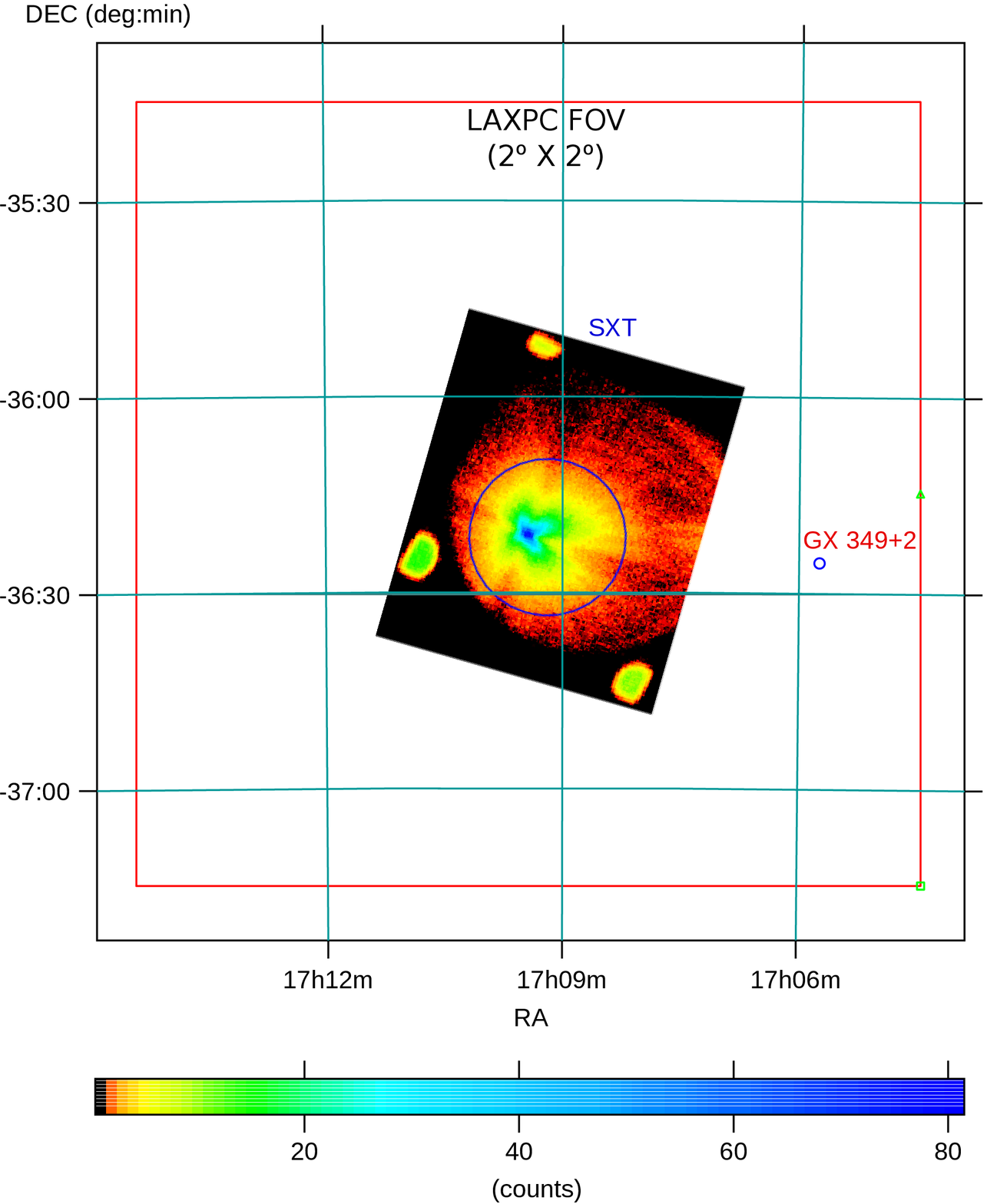}
	\includegraphics[scale=0.39]{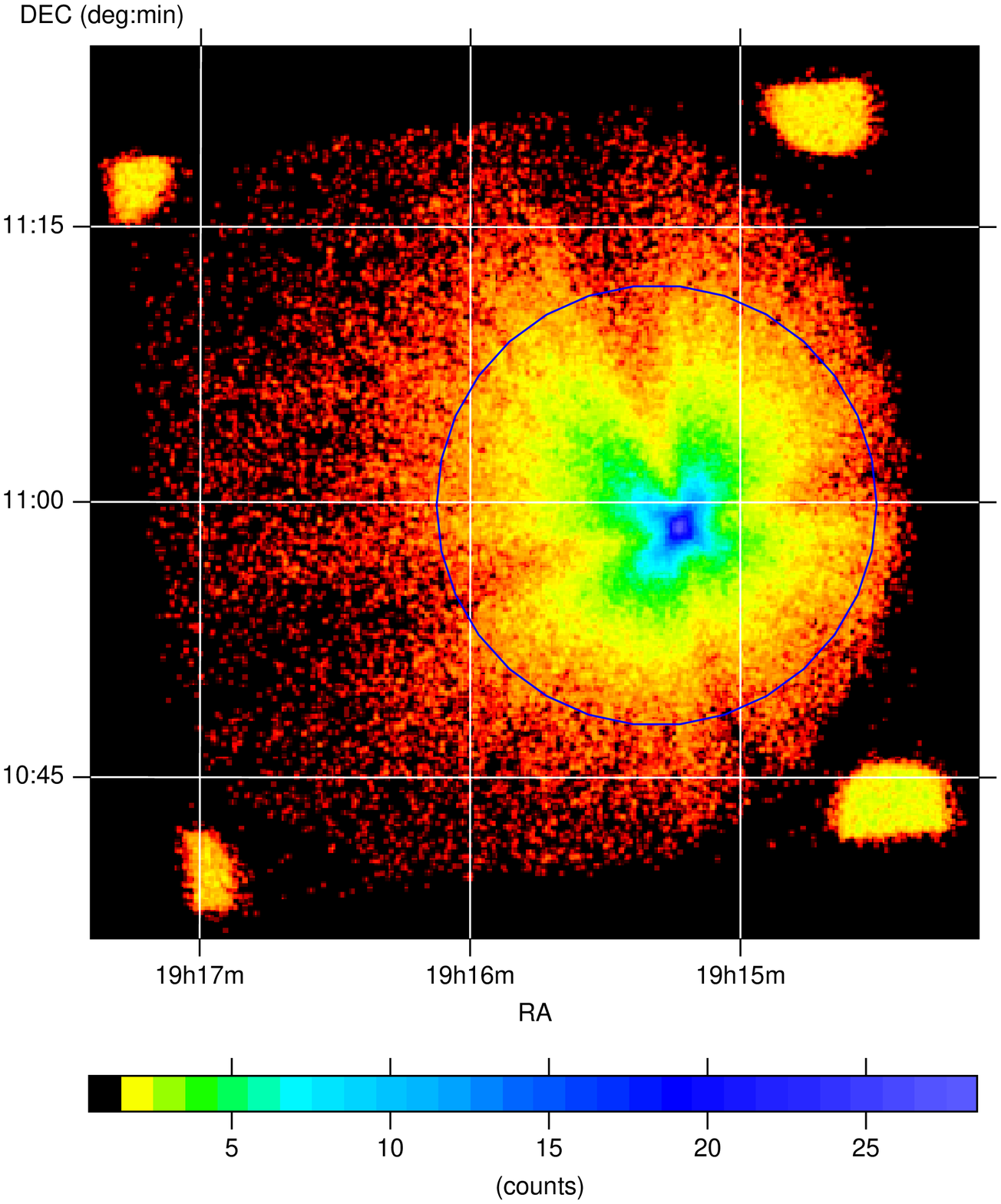}}
	\caption{IGR J17091--3624 image generated from the \textit{SXT} instrument onboard \textit{AstroSat} was rotated by 110$^{\circ}$ to align in the direction of nearby source GX 349+2 is shown in the left panel. A blue circular region of 12 arcmin radius as shown in the image was used for extracting genuine source events. The colour bar (below) provides the intensity scaling in the \textit{SXT}. A red square representing the \textit{LAXPC} FOV ($2^{\circ}\times2^{\circ}$) along with position of nearby source GX 349+2 is also marked. The right panel shows the \textit{SXT} image of GRS 1915+105 source with 12 arcmin radius (blue) for selecting the events.
    }
	\label{fig:1}
\end{figure*}

For the IGR J17091--3624, the events were selected within circular region of 12 arcmin radius (blue circle) as seen in the left panel of Fig.~\ref{fig:1}. The source image is generated for the complete exposure time of $\sim 37$ ks. The core central pixel of the source registered maximum around 83 counts. The highest count rate noticed among the individual segments is 12.79 c/s (see Table \ref{tab:1}) which is well below the pile-up limit of 40 c/s for PC mode as mentioned in the \textit{AstroSat Handbook}\footnote{\url{ https://www.issdc.gov.in/docs/as1/AstroSat-Handbook-v1.10.pdf}}.

Similarly, for the GRS 1915+105 observation, event selection was made from a circular source region with 12 arcmin radius as seen in the image on right panel of Fig.~\ref{fig:1}. The figure presented is for $\sim 4.7$ ks which is the GTI for entire duration of the `heartbeat' state as observed by \textit{AstroSat}. Of these, we choose an uninterrupted observation for 1201 s where the central pixel has a maximum $\sim 28$ counts that is below the pile-up limit of \textit{SXT}. 

\begin{figure*}
	\centering
	{
		\includegraphics[angle=-90,scale=0.33]{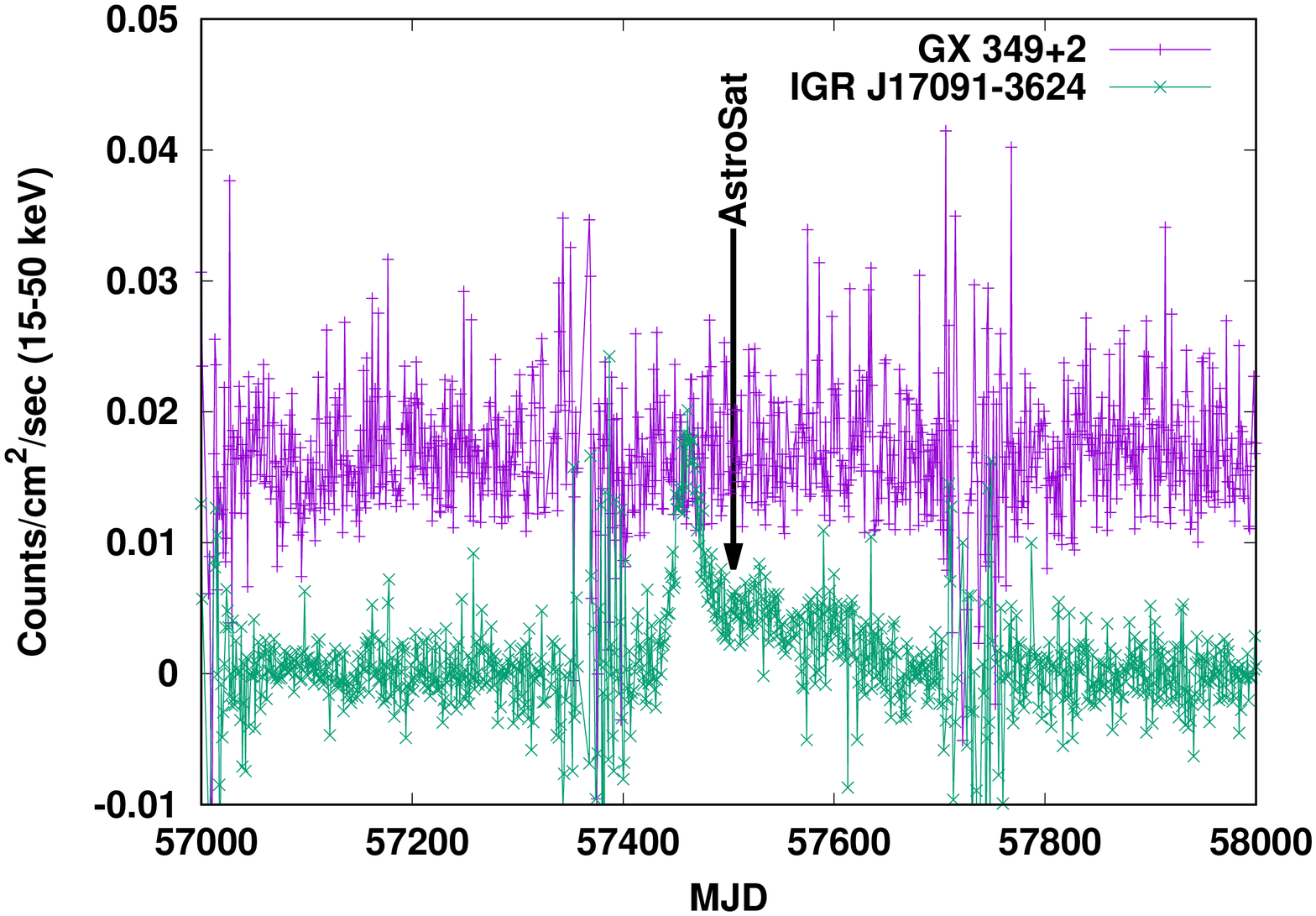}
		\includegraphics[angle=-90,scale=0.39]{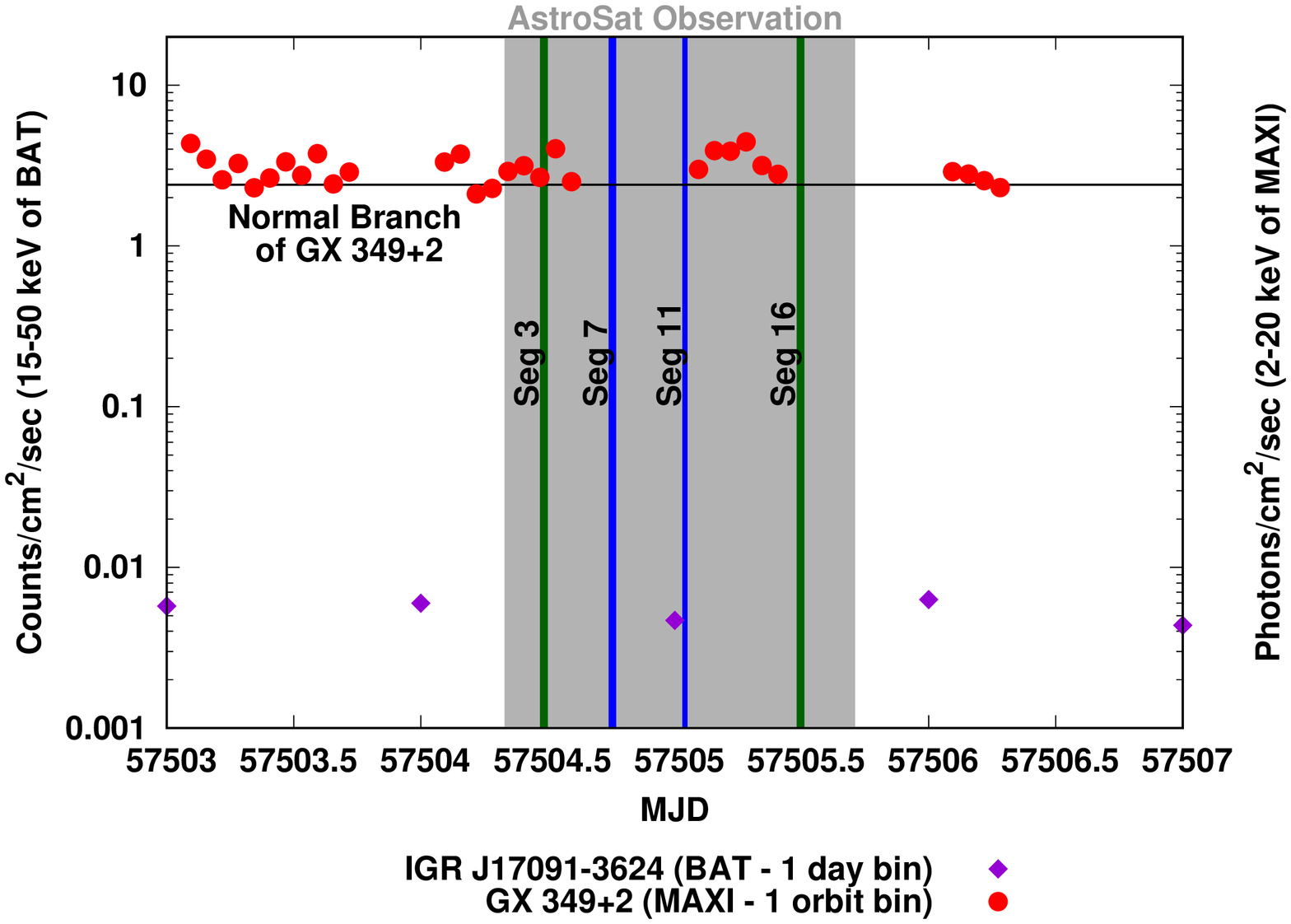}
	}
	\caption{\texttt{Left} - \textit{BAT} 1-day bin light curves of IGR J17091--3624 along with the nearby source GX 349+2 are represented by green and purple data points respectively. \textit{AstroSat} observation is marked with a black arrow. \texttt{Right} - \textit{BAT} 1-day bin light curve of IGR J17091--3624 along with the nearby source GX 349+2 1-orbit bin light curve from \textit{MAXI} are represented by purple and red data points respectively. \textit{AstroSat} observation is shown as a gray band. The black horizontal line represents the average intensity of 2.4 Photons/cm$^{2}$/sec of the Normal Branch of the source GX 349+2 during the \textit{MAXI} observation of 6 June 2016. The green and blue vertical lines shown are the selected segments of \textit{AstroSat} observation for the broadband studies, where the width of line indicates the duration of them. The green lines indicate the observation for  single peak QPO, whereas the blue lines are for the double peak QPO (see text for details).}
	\label{fig:2}
\end{figure*}

\subsection{LAXPC Data Reduction}

\textit{LAXPC} data were processed by \textit{LAXPC} pipeline software\footnote{\url{https://www.tifr.res.in/~astrosat_laxpc/LaxpcSoft.html}} (\texttt{laxpcl1} and \texttt{backshiftv3} version 3.1) released on 4 Sep 2019. We follow \cite{Antia2017,Agrawal2018,Sreehari2019,Baby2020} to obtain the light curves and energy spectrum. IGR J17091--3624 and GRS 1915+105 light curves were generated in the energy range  3--20 keV with a time-bin of 0.01 s. All the three \textit{LAXPC} units were used to generate the light curve and perform the timing analysis. 

For IGR J17091--3624 and GRS 1915+105 observation, only \textit{LAXPC20} was used to generate the spectra in the energy range $ 3-40 $ keV. To minimize the residuals due to instrumental effects in the spectra beyond 30 keV, we considered only the top layer of the detector along with the events reported as single detection of the event. However, contribution to the source spectra of IGR J17091--3624 from the nearby persistent source GX 349+2 cannot be ruled out during the observation. In the next section, we attempt to quantify and mitigate the effects of contamination in the IGR J17091--3624 source spectra.

\begin{figure}
	{\includegraphics[angle=-90,scale=0.35]{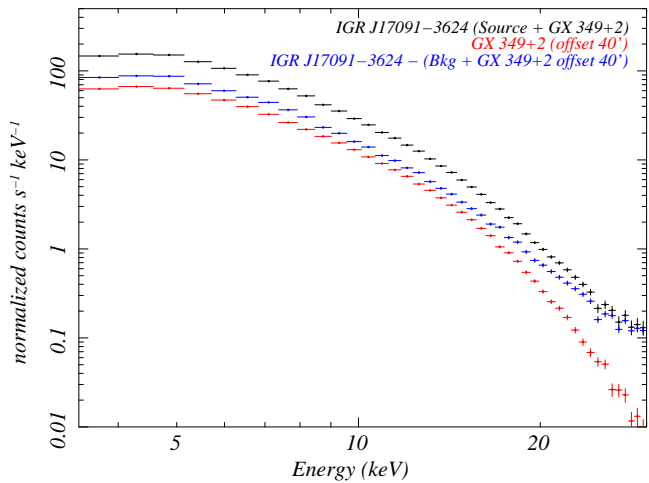}}
	\caption{\textit{LAXPC20} contaminated spectrum of IGR J17091--3624 (Source + GX 349+2) after background subtraction is shown with the black colour. The simulated spectrum (off-axis) of nearby source GX 349+2 as generated using the model \texttt{TBabs(bbodyrad + diskbb + diskline)} of the normal branch \citep{Coug2018} is plotted with the red colour. The contamination is removed from the extracted spectrum of IGR J17091--3624 after subtracting the simulated GX 349+2 spectrum, which is displayed in blue colour.}
	\label{fig:2b}
\end{figure}

\subsection{Contamination from the Nearby Source GX 349+2}
 \label{sec:cont}
The left panel of Fig.~\ref{fig:1} shows the \textit{SXT} image of IGR J17091--3624 superimposed over the \textit{LAXPC} FOV marked by the red square. 
Since the LAXPC FOV has a FWHM close to $1^\circ$, a square having the side of $2^\circ$ marks the FOV.
IGR J17091--3624 is close to the centre as seen in the \textit{SXT} image along with the nearby neutron star GX 349+2 marked at RA $17^h\,05^m$ and Dec $-36.42^{\circ}$ at an angular separation of $41\arcmin$. For a comparison of the intensity of the two sources, the left panel of Fig.~\ref{fig:2} shows the \textit{BAT} light curves of IGR J17091--3624 (green points) and GX 349+2 (purple points) in the $ 15-50 $ keV energy range. A contamination from a bright source out of the field of view seems to be present also in the SXT image. Rotation of the image by 110$^{\circ}$ aligns it in the direction of GX 349+2 source. Also, analysis of \textit{SXT} observations of GX 349+2 available at a later date was performed to confirm if contribution from the source could extend to the FOV in consideration. Since no other bright X-ray sources are reported in the direction of the source for the duration of this observation, it is reasonable to assume that the contamination in the image is stray light from GX 349+2, but, in any case, it is negligible ($\simeq1.5\%$) with respect to the IGR J17091--3624 counts. The situation is different for \textit{LAXPC} detector, in fact it is expected to have an efficiency of 30\% of the nominal value at an angular separation of $\sim40\arcmin$ from the on-axis pointing \citep{Antia2017}. Also, an additional broad variability, absent in the \textit{SXT} light curve, is seen in the \textit{LAXPC} light curve for the same duration which cannot be attributed to this source. Therefore, it is possible that the \textit{LAXPC} data for this duration was contaminated by the source GX 349+2. Due to lack of simultaneous observations by other imaging/pointed missions, we attempt to model and subtract the contribution from the source guided by previous studies of the source and using the off-axis response \citep{Antia2017,Baby2020}.
 
In order to quantify the contamination of nearby source GX 349+2 in the \textit{AstroSat} observation data, we considered the results presented by \cite{Coug2018} on the various states of GX 349+2 source as observed by the \textit{NuSTAR} observation on 6 June 2016. The corresponding MAXI 1-orbit binned light curve (red) of GX 349+2 was plotted against IGR J17091--3624 BAT 1-day binned light curve (purple) along with AstroSat observation (gray band) which are presented in the right panel of Fig.~\ref{fig:2}. We used \textit{MAXI} data for reference as it was closest to \textit{AstroSat} observation. The upper limit of flux for the normal (Non-flaring) branch of GX 349+2 was estimated using \textit{NuSTAR} observations. The corresponding flux level in the \textit{MAXI} light curve was obtained as 2.4 photons cm$^{-2}$ s$^{-1}$. We discard those observations where flux level is clearly above this base level as they could be in the flaring branch. Applying this rigid filter, we select four segments from the \textit{AstroSat} observation that are marked with blue and green vertical lines for the broadband spectral studies. The green lines indicate the segments with detection of single peak QPO and the blue line represent the segment with double peak QPO whereas the width of lines indicate the segment duration.

Although the selected observations could be less contaminated than those closer to the flaring branch of GX 349+2, it is necessary to quantify the contribution from this source to ensure that the contamination is removed. Therefore, we generate simulated spectrum for GX 349+2 using an off-axis response at 40$ \arcmin $ based on the previous studies of the source considering it to be in the normal branch. The simulated spectrum is generated with a $0.1-200$ keV flux of $2.39\times10^{-8}$ ergs cm$^{-2}$ s$^{-1}$ \citep{Coug2018} using the off-axis response. Comparison with the source spectrum obtained from \textit{LAXPC} is shown in the Fig.~\ref{fig:2b}. The model used to generate the simulated spectrum is \texttt{TBabs(bbodyrad + diskbb + diskline)}. It can be seen that the contribution from GX 349+2 at an offset of $40\arcmin$ (red) cannot be considered negligible. We therefore, consider this simulated spectrum as an additional background in \texttt{XSPEC} during spectral analysis which gives a spectrum with significant counts up to 30 keV. However, the contribution from GX 349+2 becomes significant beyond 23 keV as seen from Fig. 3. Hence, spectral analysis is performed only up to 23 keV, to rule out modification of spectral parameters due to contribution by GX 349+2 at higher energies (see, Sect.~\ref{sec:sm} \& Table~\ref{tab:4}).
 
\begin{figure*}
	{\includegraphics[width=8.5cm,height=6.5cm]{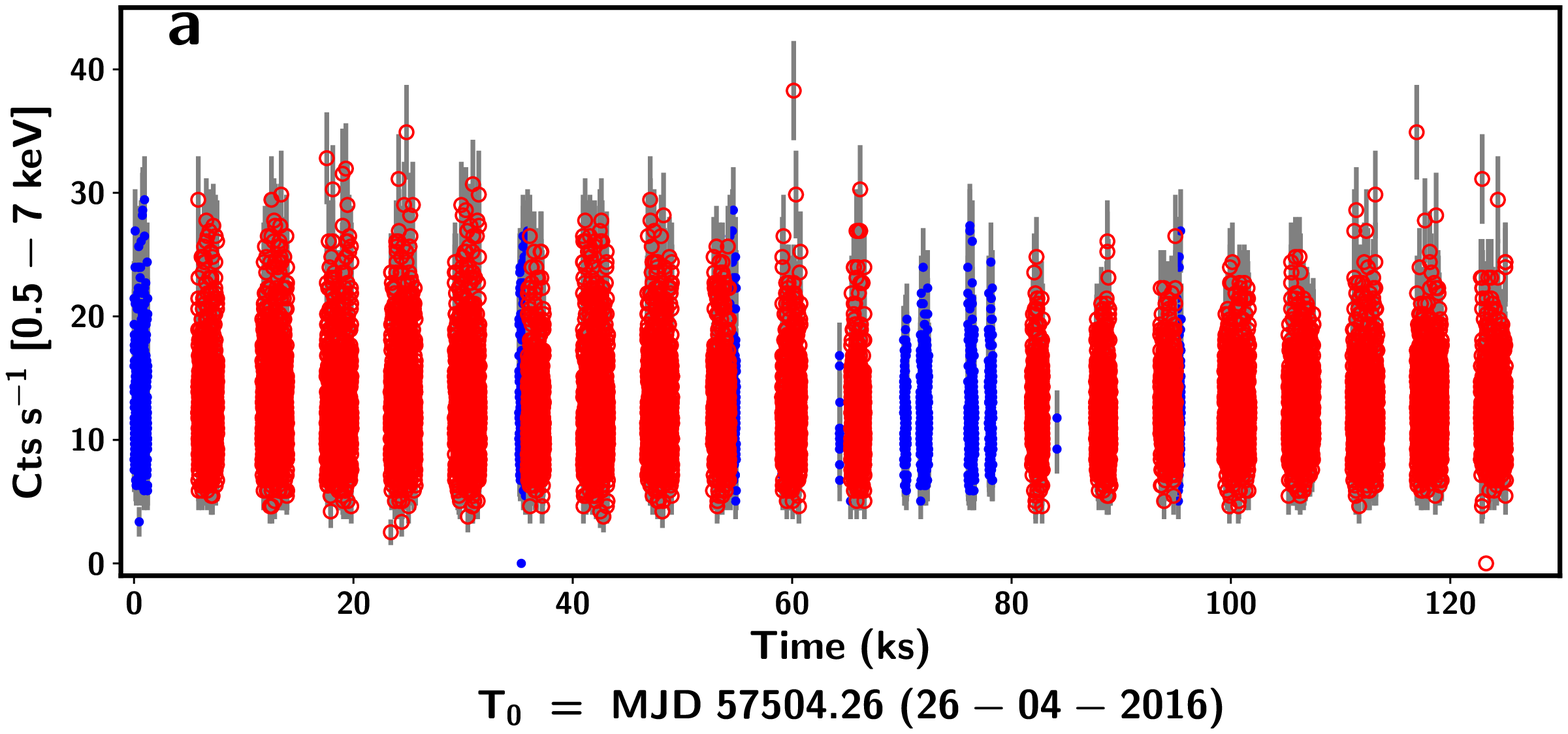} 
	 \includegraphics[width=8.5cm,height=6.9cm]{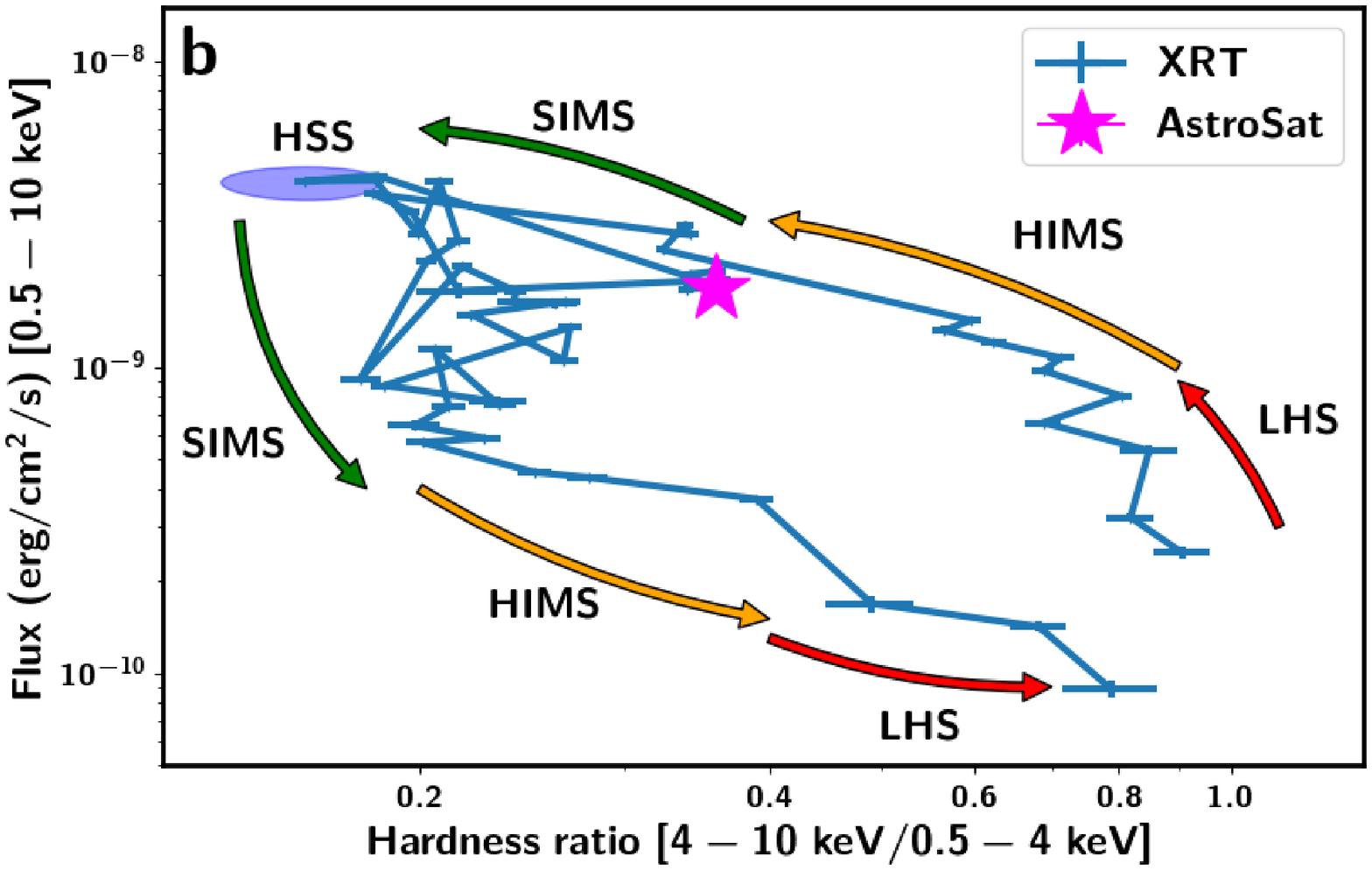}}
	\caption{(a) The \textit{SXT} light curve of the source IGR J17091--3624 during its 2016 outburst. The selected 19 segments ($\ge1$ ks) are marked with red colour. The blue coloured segments are not chosen either due to an exposure time $\le1$ ks or presence of data gap of the order of tens or hundreds of seconds between them. (b) HID showing evolution of the hardness ratio with flux is plotted using \textit{Swift/XRT} observation during the 2016 outburst. The \textit{AstroSat} observation is marked with a magenta star which is in the SIMS \citep[see][for details]{Radhi2018} of the 2016 outburst.}
    \label{fig:3}
\end{figure*}
 
\section{Analysis \& Modelling}
\label{ana}
\textit{AstroSat} provides us the opportunity to study the nature of IGR J17091--3624 in the broadband energy range during its 2016 outburst. However, due to the limitation mentioned in Sect.~\ref{sec:cont}, we used the \textit{LAXPC} data from  $3.0-23$ keV for spectral analysis on selected observations and $3-20$ keV timing from \textit{LAXPC} is done to emphasize the timing results from \textit{SXT}. For GRS 1915+105, energy range of 0.7--7 keV for \textit{SXT} and 3--40 keV for \textit{LAXPC} respectively were considered for temporal and spectral analysis. 

In the following subsections, we summarize the methodologies used to analyse temporal and spectral data for both these instruments. 

\subsection{Temporal Modelling}
\textit{SXT} light curve is plotted in Fig.~\ref{fig:3}a. The Hardness-Intensity Diagram (HID) is the variation of intensity with the ratio of flux in the  harder to softer band. It is a diagnostic tool with which to classify different states during an outburst \citep{Homa2005, Bell2005, Nandi2012, Nandi2018, Sreehari2019}. The HID obtained from \textit{Swift/XRT} is shown in Fig.~\ref{fig:3}b. The \textit{AstroSat} observation is marked as a star which lies in the SIMS region \citep[see][]{Radhi2018}. The light curves in 0.5--7 keV energy band with a time bin of 2.378 s was used to create a power spectrum of the selected segments. The \textit{powspec} tool of \texttt{XRONOS} version 5.22 was used. A total of 256 bins per interval and 3 intervals in a frame were considered for obtaining the power spectrum in the units of (rms/mean)$^2$/Hz. We also applied a geometric re-binning factor of 1.05. We follow this approach to fit all the PDS with multiple Lorentzian functions to model `heartbeat' QPOs, its harmonics and with a power-law component whenever required. The significance of `heartbeat' QPOs is estimated as the ratio between the normalization and its negative error \citep[see][and references therein]{Sreehari2019}.

\begin{figure*}
	\includegraphics[scale=0.40]{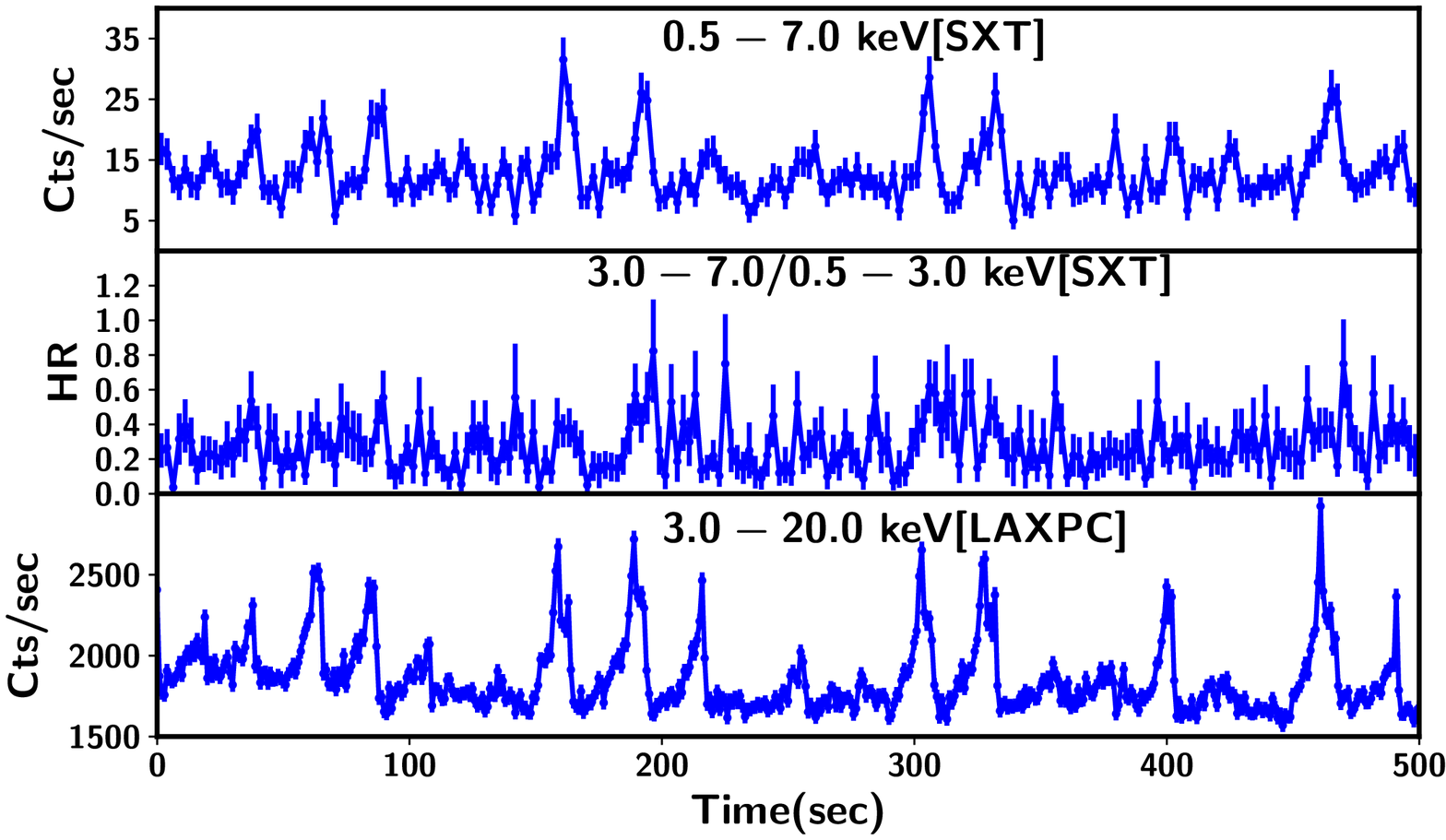}
	\includegraphics[scale=0.345]{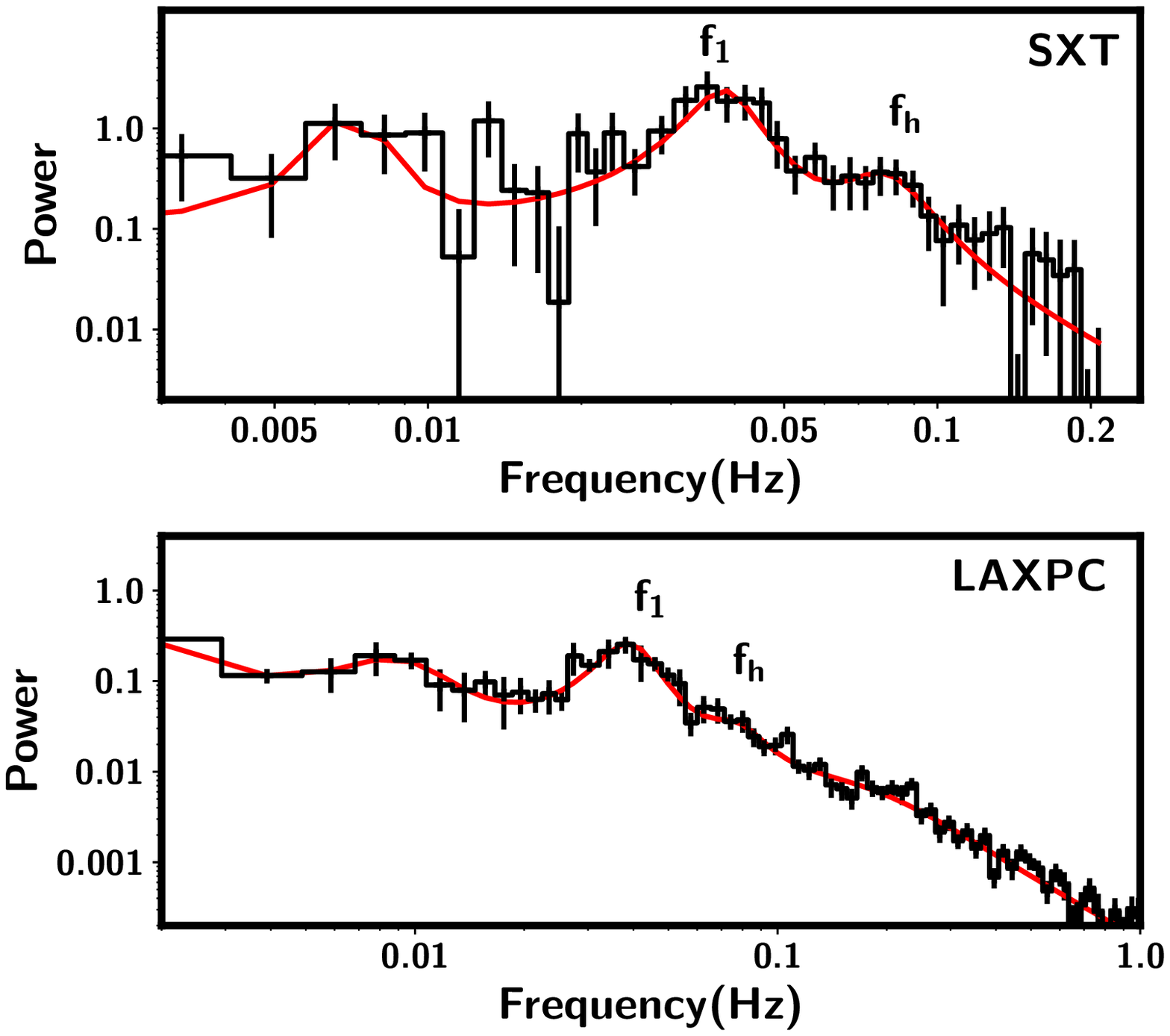}
   	\caption{ \texttt{Left} - Top panel of figure displays the \textit{SXT} light curve for Segment 3, along with HR (3--7 keV/0.5--3 keV) in the middle panel. \textit{LAXPC} light curve is plotted for the same observation time in the bottom panel which shows the burst profiles similar to \textit{SXT}. \texttt{Right} - The top and bottom panels of the figure show the corresponding PDS from \textit{SXT} (top) and \textit{LAXPC} (bottom) data respectively. Single peak `heartbeat' QPO is seen in both PDS around 38.9 mHz. Data from the last 500 s of Segment 3 is plotted in the light curve for clarity, whereas the whole segment is used to generate the PDS.}
   	\label{fig:4}
\end{figure*}

The average power spectra generated for Segments 3 and 4 of \textit{SXT} are shown in Figs.~\ref{fig:4} \& \ref{fig:5} (right panels). Initially, the power spectrum of Segment 3 is fitted with a Lorentzian for the fundamental `heartbeat' QPO which results into a $\chi^2_\mathrm{red} = 1.41~(56.23/40)$. However, with the inclusion of two more Lorentzian features, the fit improved significantly with a $\chi^2_\mathrm{red} = 0.87~(29.53/34)$. Similarly, we model the power spectrum for Segment 4 with multiple Lorentzian features, considering the fundamental `heartbeat' QPO feature and harmonic. The fitted result yields a $\chi^2_\mathrm{red} = 1.67~(51.64/31)$. The fit significantly improved after inclusion of an additional Lorentzian for the second `heartbeat' QPO which results into a $\chi^2_\mathrm{red} = 1.16~(32.45/28)$. Such double peaks were observed in seven of the 19 segments considered at mHz frequencies (see Table \ref{tab:2} and Figs.~\ref{fig:4} \& \ref{fig:5}).

Further, we studied the dynamic PDS using the Interactive Data language (IDL) version 8.5. Using the IDL routine \textit{WV\_CWT} (continuous wavelet transform) for wavelet analysis, dynamic PDS were generated for these burst profiles for Segments 3 and 4. This routine transforms the time series data to power spectrum by using the \textit{Morlet} wavelet functions consisting of a complex exponential modulated by a Gaussian envelope. For the wavelet analysis, the function parameters were set to \textit{Morlet} family and \textit{order} to 6 (width of wavelet). The main purpose of using the dynamic power spectrum was to check if the double peak in some segment is due to variation in QPO frequency during the observation.

Similarly, the temporal analysis of \textit{LAXPC} data was carried out for the energy range $ 3-20 $ keV. The light curves of all the units were generated with time resolution of 0.01 s and were merged together, however, the PDS was generated for the time resolution of 0.5 s with 1024 intervals. Thus, PDS plot has a frequency range of 2 mHz -- 1 Hz which is normalized to obtain the squared fractional rms variability. `Heartbeat' QPOs were also observed in the \textit{LAXPC} PDS at similar frequencies. Double peak frequencies profile, similar to those observed in \textit{SXT} PDS were also found in the \textit{LAXPC} PDS for the same segments which confirms their `true' detection (see Figs.~\ref{fig:4} \& \ref{fig:5}).

\textit{SXT} PDS was generated for the same frequency range in the `heartbeat' state of GRS 1915+105, whereas PDS of  \textit{LAXPC} data was generated with Nyquist frequency of 10 Hz, using time resolution of 0.05 s with 8192 intervals. It is normalized to obtain the squared fractional rms variability with appropriate dead-time correction applied. Both PDS are fitted with multiple Lorentzians and a power law function.

\subsection{Spectral Modelling}
\label{sec:sm}
The spectral analysis and modelling was carried out using \texttt{XSPEC v12.10} tool of \texttt{HEASOFT v6.26} for the \textit{SXT} data in the energy range 0.7--7 keV. First, we model the IGR J17091--3624 spectrum with an absorbed multi-colour disc model \texttt{diskbb} \citep{Mitsuda1984,Makishima1986}. Interstellar absorption was accounted for by \texttt{TBabs} model. Acceptable fits were obtained with $\chi^2_\mathrm{red} = 1.14~(467/410)$. As the source was in SIMS, where the disc extends closer to the compact object, we replace \texttt{diskbb} with \texttt{ezdiskbb} \citep{Zimm2005} which considers zero-torque condition at inner disc boundary with marginal improvement in the fit. The model used was \texttt{TBabs*ezdiskbb} (hereafter Model 1). Further, we extended this model with broadband spectral analysis using both \textit{SXT} and contamination free \textit{LAXPC} spectral data (by subtracting the simulated GX 349+2 spectrum) in the 0.7--23 keV range by considering the model \texttt{simpl*ezdiskbb+gaussian} to account for the Comptonisation of disc photons by the optically thin corona \textbf{and to investigate the presence of iron line.} However, the inclusion of \texttt{simpl} made gain correction non-executable over \textit{SXT} data. This led to poor fits with the instrumental edges at 1.8 and 2.4 keV being unaccounted for. Also, significant residuals were seen in the $10-20$ keV range. Thereafter, \texttt{nthcomp}, which is an advanced model, was applied to describe the Comptonisation component in the hard energy band. The input seed photon temperature of \texttt{nthcomp} was tied up with disc temperature of \texttt{ezdiskbb} model. Inclusion of a \texttt{gaussian} line at $\sim$ 6.4 keV resulted in better fits. The final model used was Model 2a: \texttt{TBabs(ezdiskbb + gaussian + nthComp)}.

The broadband spectral analysis of GRS 1915+105 was performed using both \textit{SXT} and \textit{LAXPC} data in the 0.5--40 keV range. This broadband spectrum required an additional \texttt{Gaussian} line and the \texttt{nthcomp} model along with \texttt{ezdiskbb} to model the higher energy part. \texttt{Gaussian} was replaced by \texttt{diskline} model which considers line emission from a relativistic accretion disc \citep{Fabian1989}. \textbf{Residuals corresponding to an absorption edge are seen in the $7-9$ keV range \citep[see][]{Done2004,Sreehari2020}. This is modelled using the smeared edge model, \texttt{smedge} in \texttt{XSPEC} \citep{Ebisawa1994}, with index for photoelectric cross-section fixed at -2.67. This improved the fit slightly with $\chi^{2}_{red}$ decreasing from 1.20 (571/474) to 1.18 (560/473). The edge is obtained at $\sim$ 8 keV with the width ranging from $2-3$ keV. The smearing width is therefore frozen at 3 keV for analysis.} The final model used was \texttt{TBabs(ezdiskbb + diskline + nthComp)} which will henceforth be referred to as Model 2b.

A systematic error of 1.5\% was incorporated for the combined fit for both \textit{SXT} and \textit{LAXPC} data. A normalization constant is considered to account for the difference in calibration of both instruments which turns out to be close to 1.

\section{Results}
\label{res}

We present the results of temporal and spectral data analysis of IGR J17091--3624 during its 2016 outburst. In Sect.~\ref{sec:hid}, we outline the evolution of the source during the outburst and categorize the current observation with regards to its spectral state. In Sect.~\ref{sec:timing}, \ref{sec:spec} and \ref{sec:phas}, we present the detailed results of timing and spectral analysis along with simple Phase-Resolved Spectroscopy (PRS) respectively, whereas in Sect.~\ref{sec:comp} we draw parallels with the `heartbeat' state of GRS 1915+105.

\subsection{Light curve and Hardness Intensity Diagram (HID) of IGR J17091--3624}
 \label{sec:hid}
\begin{figure*}
	\includegraphics[scale=0.41]{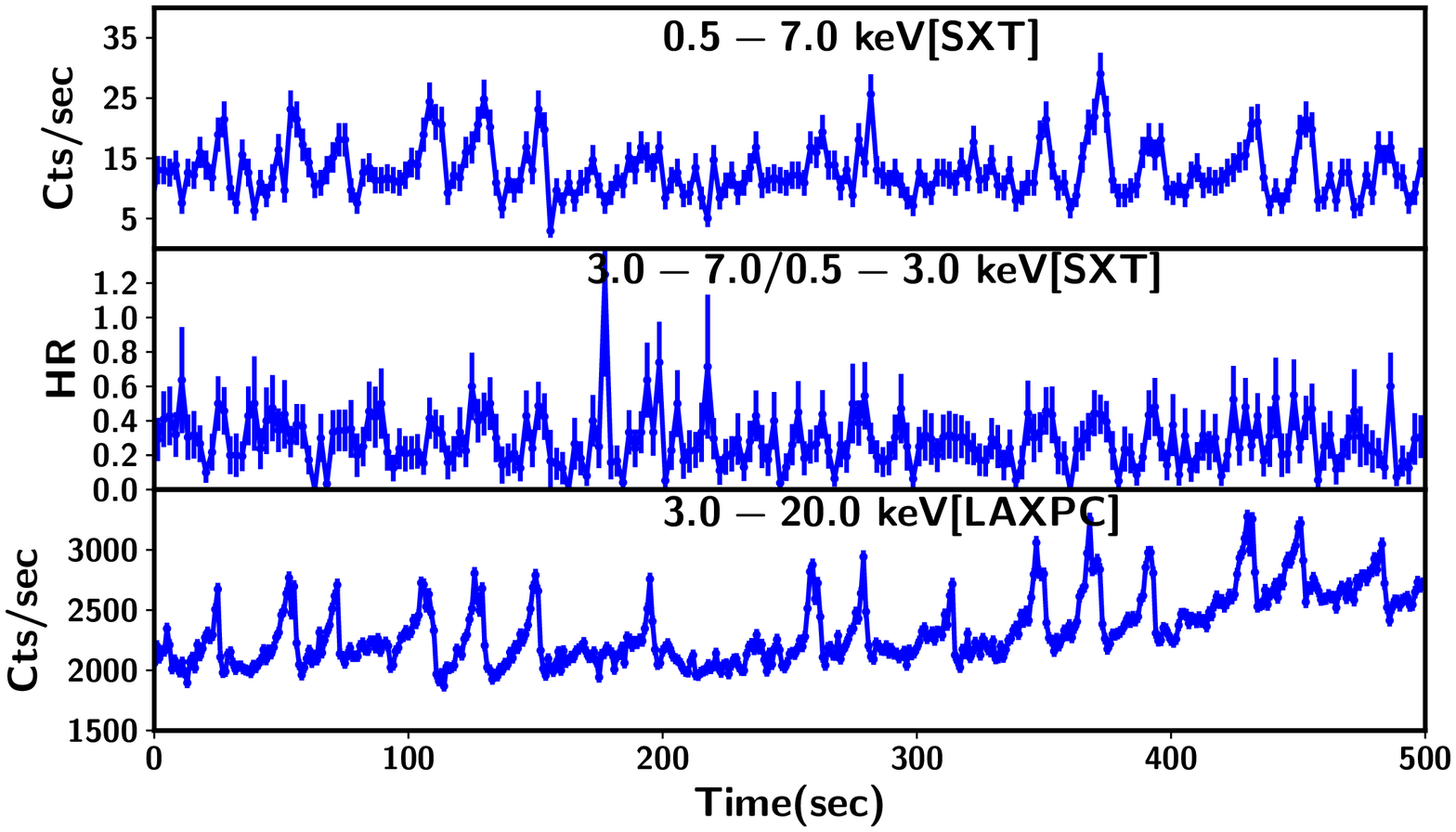}
	\includegraphics[scale=0.36]{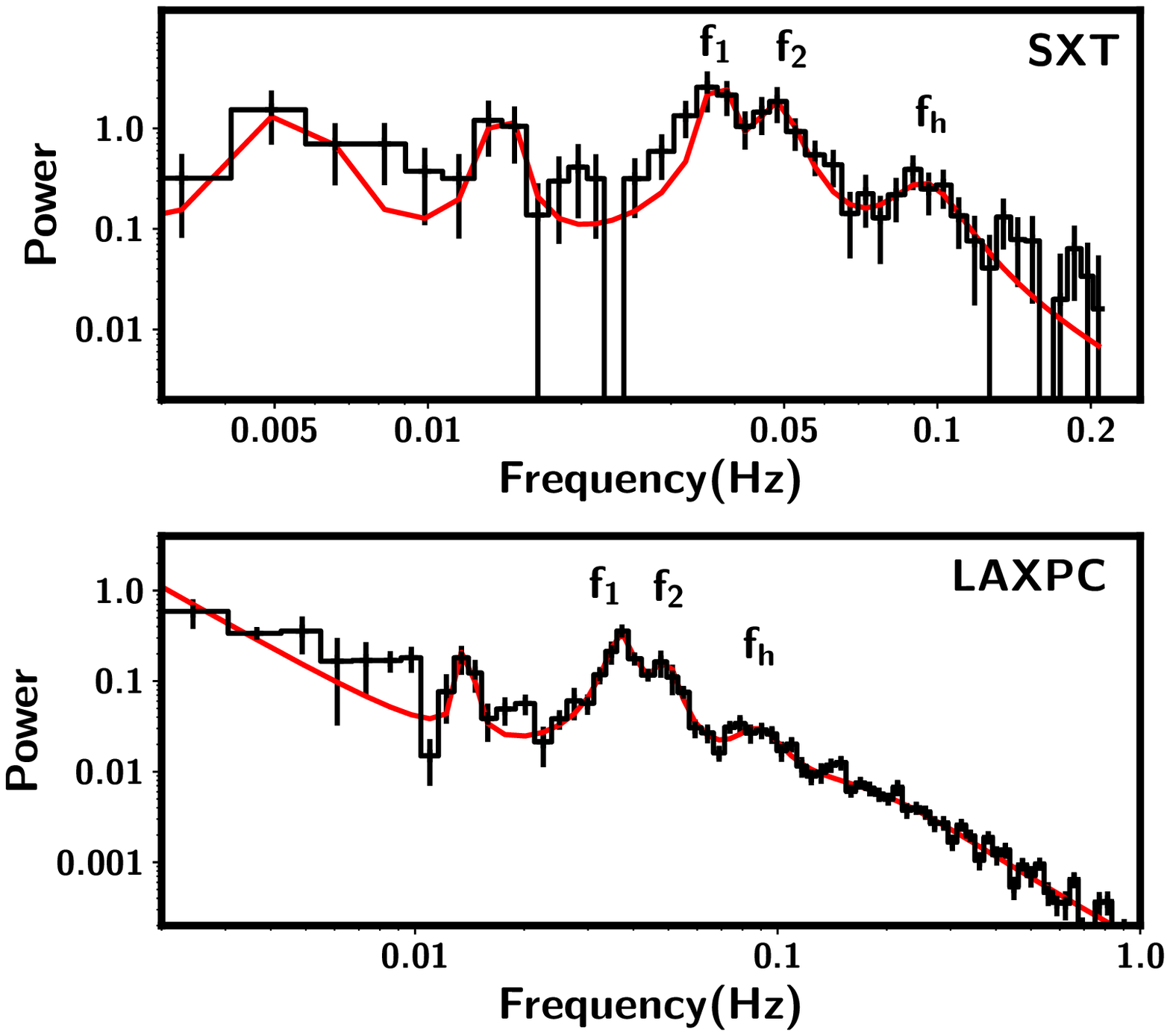}
	\caption{Left - The presentation of the figure for Segment 4 follows the same convention as in Fig.~\ref{fig:4}. Right - Double peaked burst frequencies at $36.8 \pm 0.44$ mHz and $48.4 \pm 1.45$ mHz are seen in the \textit{SXT} PDS (top panel). In the bottom panel, for the same observation time, the \textit{LAXPC} PDS also showed double peaked burst profile at the frequency of $37.1 \pm 0.54$ mHz and $47.8 \pm 0.65$ mHz, similar to the \textit{SXT} PDS.}
	\label{fig:5}
\end{figure*}

 \textit{SXT} light curve of the entire IGR J17091--3624 observation with time bin size of 2.378 s is shown in Fig.~\ref{fig:3}a. The source intensity is observed to vary between 5--35 c/s in the energy band 0.5--7 keV. The upper left panels of Fig.~\ref{fig:4} \& \ref{fig:5} show the 500 s zoomed \textit{SXT} light curve of Segments 3 and 4 in the energy band 0.5--7 keV, whereas the middle panel shows the variation of hardness ratio (HR) with time where HR is defined as ratio between flux in 3--7 keV and 0.5--3 keV. We also plot the \textit{LAXPC} light curve in the energy band 3--20 keV in the bottom panels and find similar variability as in \textit{SXT} light curve. A similar type of variability was observed during the `heartbeat' phase of the 2011 and 2016 outbursts \citep{Capi2012, Radhi2018} as observed with \textit{Swift/XRT} and \textit{NuSTAR}. The source is in the soft-intermediate state considering the low hardness ratios as was also stated by \cite{Radhi2018}. Lower HR values were also obtained for the \textit{LAXPC} light curve, however the data are not presented here due to possible contamination by the neutron star GX 349+2 source. The evolution of HID in terms of hardness ratio (flux in 4--10 keV/0.5--4 keV) and total unabsorbed flux in 0.5--10 keV (as observed with \textit{Swift/XRT}) is shown in Fig.~\ref{fig:3}b \citep[see][for details]{Radhi2018}.

\begin{figure*}
	\includegraphics[scale=0.1]{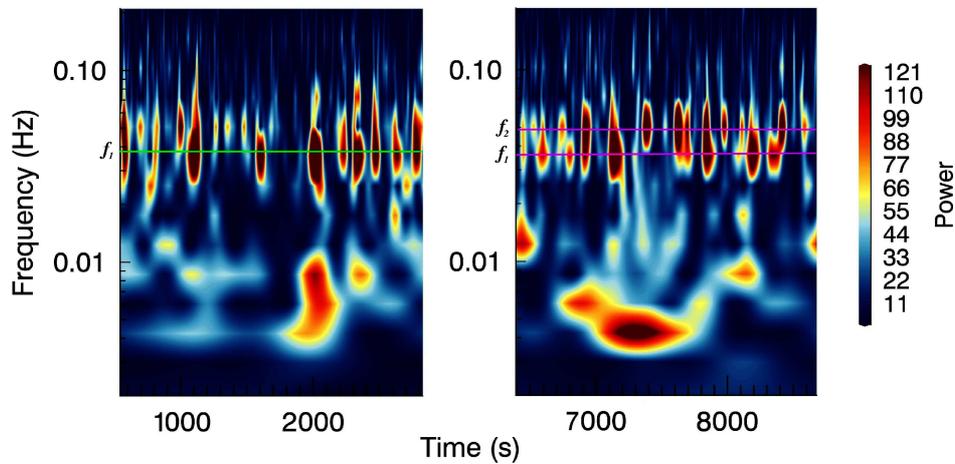}
	\caption{Wavelet analysis of \textit{SXT} light curve showing the Dynamic PDS for Segments 3 and 4. Segment 3 (left, the frequency $f_{1}$ marked with green line) shows only single burst frequency (see Fig.~\ref{fig:4}), whereas Segment 4 (right, frequencies $f_{1}$ and $f_{2}$ marked with violet lines) shows the double burst frequency (see Fig.~\ref{fig:5}). Clear evolution from single peak (Segment 3) to double peak (Segment 4) is seen in the dynamic PDS (see text for details).}
	\label{fig:6}
\end{figure*}

\subsection{Timing Properties}
 \label{sec:timing}

Evolution of temporal properties in segments, where single peaked and double peaked QPOs are seen is described using successive segments 3 and 4 as a representative case for clarity. \textit{SXT} light curves of segments 3 and 4 are shown in Fig. \ref{fig:4} and \ref{fig:5} where structured variability is clearly visible. We observe two distinct types of peaks (`heartbeat' QPOs) in the power spectra i.e.,  single and double peaks along with the harmonics. The PDS is fitted with power law and Lorentzian features for single peak frequency ($f_{1}$) in the range of $ 25.6-40.7 $ mHz along with harmonics observed at 54.0--87.2 mHz ($f_{h}$). The power spectra also show an additional peak ($f_{2}$) in some of segments between 34.7--48.4 mHz. A single main frequency $f_{1}$ was detected in almost all the selected segments. The rms of `heartbeat' QPO ($f_{1}$) were obtained between 10.3--22.4 \%, whereas total rms lies in the range of 16.5--28.6 \%. However, rms of the harmonics was found to be in the range of 3.5--14.7 \% and the significance of peaks was found to be small (see Table \ref{tab:2}). It is also observed that the frequency of `heartbeat' QPO decreased from Segment 11 onwards to a frequency in 25.6--30.3 mHz, as compared to the 33.1--40.7 mHz in the initial Segments (see Table \ref{tab:2}). Apart from these, a distinguishable low frequency feature between 5.5--10.0 mHz is also seen along with its harmonic between 10--20 mHz in some cases. The harmonic feature could be discerned in a few other cases where the main peak could not be obtained. Similar features and frequency ranges were also observed with \textit{LAXPC} PDS as presented in Table \ref{tab:2}.

In the following sections, we focus on the temporal analysis considering the single peak and double peak frequency structure in the PDS for Segments 3 \& 4 (as representation for two independent cases). 

\subsubsection{Single Peaked QPO (SPQ)}
 \label{sfp}

The single peaked `heartbeat' QPO observed in segment 3 is shown in Fig.~\ref{fig:4}. The burst oscillation frequency shows up as a feature at 38.5 mHz ($f_{1}$) for Segment 3. A harmonic was also noticed at 78.6 mHz ($f_{h}$). An additional broad feature was seen at 7.2 mHz (see Fig.~\ref{fig:4}, top--right panel). The total rms of the fitted model over the frequency range considered was 27.1\% of which the `heartbeat' QPO contributed 20.2\%. The rms of harmonic was obtained at half of this value. 

In order to confirm the frequency obtained using the \textit{SXT} light curve for this segment, the simultaneous \textit{LAXPC} PDS is also generated as seen in Fig.~\ref{fig:4} (right--bottom panel). This also shows the frequency of `heartbeat' QPO around 38.4 mHz ($f_{1}$) and harmonic at 76.0 mHz ($f_{h}$) along with the broad feature at 8.5 mHz. However, the total rms values are not quoted due to contamination by the nearby source GX 349+2 as mentioned in Sect \ref{sec:cont}.

This SPQ is observed in 12 of the 19 segments albeit with less power in some of the segments as presented in Table \ref{tab:2}.

\subsubsection{Double Peaked QPO (DPQ)}
 \label{dfp}

As seen in the PDS of Segment 4 of \textit{SXT} data, along with the features seen in Segment 3, an additional Lorentzian was required at 48.4 Hz ($f_2$) which is shown in Fig.~\ref{fig:5} (right--top panel). It is evident that the frequencies $f_1$ and $f_2$ are not harmonically related with a separation of 11.6 mHz between them (see Fig.~\ref{fig:5} and Table \ref{tab:2}). However, only one harmonic peak is detected at 94.4 mHz ($f_h$). The total rms of the fitted model was 27.2\% with the main frequency ($f_{1}$) contributing 16.2\% and $f_2$ contributing 15.4\%. The rms estimated for the harmonic was 9.6\%. We also confirmed the presence of DPQ in the \textit{LAXPC} PDS as well (right--bottom panel). Two distinct peaks were obtained at 37.0 mHz ($f_1$) and 49.1 mHz ($f_2$) close to the frequencies observed in \textit{SXT} PDS with separation of 12.1 mHz between them. A single harmonic peak was also present at 88.2 mHz ($f_h$). Model fitted parameter of PDS of DPQ are presented in Table \ref{tab:2}. 

\subsubsection{Evolution from SPQ to DPQ}

SPQ and DPQ observed in Segments 3 and 4 respectively are separated by $\sim 4$ ks. In order to verify that the double peak in some segments is due to variation in QPO frequency during the observation, a dynamical power spectrum using the wavelet analysis was generated using the IDL tool and the results are shown in Fig.~\ref{fig:6}. The dynamic PDS gives an impression that these bursts are not continuous but rather the source is quiet for some duration in between the bursts. In Segment 3 (left hand side of Fig.~\ref{fig:6}), the SPQ is clearly observed in the dynamic PDS. They appear nine times almost consistently around 39 mHz ($f_{1}$) at the same frequency as that in the static PDS, with a wavelet power signal contour having broad width distribution within $ \pm10$ mHz in the interval of 2.3 ksec. Harmonics ($f_{h}$) at 78 mHz are also noticed for some duration.
A very faint signal is also noticed at 50 mHz near beginning of the segment. A broad wavelet signal around 9 mHz low frequency, which is widely spread over, is detected at the mid-segment with a harmonic signal at 18 mHz which is also seen in the static PDS (Fig.~\ref{fig:4}.)  

As seen in Fig.~\ref{fig:6} (right hand side), these burst frequencies show a split by swapping in the frequency space due to shift in their location with time in Segment 4. Initially, we observed three burst signals which are weak in strength resembling a jump in frequency space resulting in a double peak. These DPQs become stronger with time as seen from the Fig.~\ref{fig:6}. Four broad burst frequencies ($f_{1}$) at 37 mHz with $\pm10$ mHz in comparison to five narrow burst frequencies ($f_{2}$) at 48 mHz with $\pm 5$ mHz are seen in the dynamic PDS. A faint signal of harmonic of frequency ($f_{h}$) around 100 mHz is noticed at a few locations in frequency space. Moreover, a strong broad signal appears around 4 mHz and a couple of weak signals at 6 mHz and 8 mHz along with harmonic signal present around 14 mHz (at the start of segment) are well noticeable. 

\subsection{Spectral Properties}
\label{sec:spec}
From the spectral fits of all the data sets, the hydrogen column density ($N_H$) is found to be (0.8--0.9)$\times 10^{22}$ cm$^{-2}$. These are similar to the estimates made by \cite{Krim2011} and \cite{Capi2012}. The \textit{SXT} spectrum is well fitted by Model 1 as shown in Fig. 8. The disc temperatures were obtained in the range 1.21--1.31 keV and the estimated flux for 2--10 keV was in the range of 1.10--1.24~$\times  10^{-9}$ ergs cm$^{-2}$ s$^{-1}$. Table \ref{tab:3} summarizes results of best fitted parameters of \textit{SXT} spectra with Model 1. No significant variation is seen in the spectra with time.

The extended broadband spectral analysis was performed on four segments with Model 2a, two of them are SPQ (Seg 03 and 16) and other two DPQ (Seg 07 and 11) segments. The fit parameters are shown in Table \ref{tab:4}. The spectrum and unfolded model for segment 7 is shown in  Fig.~\ref{fig:lxpspc}. The $N_{H}$ was kept frozen to the value obtained from Model 1, as it is better estimated from the soft energy spectrum. The disc was considered to be the source of the input seed photons in the \texttt{nthcomp} model. The \texttt{ezdiskbb} model \textit{norm} was also frozen to the value obtained from Model 1, in order to achieve better estimation of the model parameters, as it was not well constrained. The temperature of the disc was found to be in the range of 0.63--0.84 keV, whereas the electron temperature remains in the range 3.51--4.02 keV. A slight variation in the photon index was seen having the value of 2.13--2.40 along with marginal shift in the iron line energy from 6.62--6.79 keV. The Fe-line is detected only in the \textit{LAXPC} spectrum. But the feature can be non-reliable due to the possible contribution from the GX 349+2 source. Lack of knowledge on the exact nature of GX 349+2 source during this observation can contribute to the residuals in this energy range. Hence, although the changes in line energy and width in source spectrum obtained post subtraction of the simulated spectrum are significantly small with respect to individual segments, we refrain from a detailed interpretation of the same. Since the spectral parameters in both cases remain similar, we further probe into the differences between the SPQ and DPQ segments using a simple PRS in which phases are classified based on flux relative to the peak in the next section.

\begin{table*}
	\setlength{\tabcolsep}{5pt}
	\centering
	\caption{Best fitted timing parameters of the \textit{SXT} and \textit{LAXPC} data of IGR J17091--3624. The errors are estimated with 90\% confidence range for each parameter.}
	\label{tab:2}
	\begin{tabular}{cccrrccccrcc} 
		\hline
		Seg. & Freq. & `Heartbeat' & \multicolumn{5}{c}{\underline{SXT}} & `Heartbeat' & \multicolumn{3}{c}{\underline{LAXPC}} \\
		(No) &    & QPO (mHz) & Q$_{fact}$ &  QPO$_{rms}$(\%) &  Sign.($\sigma$) & Total$_{rms}$(\%) & $\chi^{2}/dof$ & QPO (mHz) & Q$_{fact}$ & QPO$_{rms}$(\%) & Sign.($\sigma$)\\
		\hline
01 & $f_{1}$ & $36.9 \pm 1.5$ &  2.9 & $18.8 \pm 3.0$ & 1.7 & $25.6 \pm 4.2$ & 25.0/32 & $36.7 \pm 1.3$ & 2.4 & $6.3 \pm 1.0$ & 1.4\\
02 & $f_{1}$ & $34.4 \pm 0.8$ & 10.7 & $13.9 \pm 3.1$ & 2.1 & $27.1 \pm 5.2$ & 28.9/31 & $35.3 \pm 0.7$ & 7.0 & $3.8 \pm 0.7$ & 1.8\\
   & $f_{2}$ & $44.1 \pm 1.3$ &  5.6 & $16.4 \pm 3.7$ & 1.9 &  -             & -       & $44.4 \pm 0.6$ & 7.8 & $5.3 \pm 1.0$ & 1.7\\
   & $f_{h}$ & $86.6 \pm 3.0$ &  2.8 & $ 8.7 \pm 2.0$ & 2.7 &  -             & -       & $79.7 \pm 1.9$ & 4.7 & $3.7 \pm 0.7$ & 1.5\\
03 & $f_{1}$ & $38.5 \pm 1.3$ &  3.3 & $20.2 \pm 3.6$ & 1.6 & $27.1 \pm 4.5$ & 29.5/34 & $38.4 \pm 0.8$ & 2.5 & $7.2 \pm 1.1$ & 1.3\\
   & $f_{h}$ & $78.6 \pm 5.0$ &  2.5 & $ 9.5 \pm 2.1$ & 1.8 &  -             & -       & $76.0 \pm 5.0$ & 2.5 & $3.8 \pm 0.7$ & 2.0\\
04 & $f_{1}$ & $36.8 \pm 0.4$ &  5.5 & $16.2 \pm 3.3$ & 2.5 & $27.2 \pm 5.1$ & 32.5/28 & $37.0 \pm 0.4$ & 6.0 & $6.2 \pm 1.1$ & 1.4\\
   & $f_{2}$ & $48.4 \pm 1.4$ &  4.4 & $15.4 \pm 3.1$ & 2.3 & -              & -       & $49.1 \pm 0.8$ & 6.3 & $4.8 \pm 1.0$ & 1.8\\
   & $f_{h}$ & $94.4 \pm 3.8$ &  3.0 & $ 9.6 \pm 2.0$ & 1.6 & -              & -       & $88.2 \pm 2.6$ & 3.0 & $3.5 \pm 0.7$ & 1.6\\
05 & $f_{1}$ & $40.7 \pm 1.3$ &  3.1 & $22.4 \pm 3.5$ & 1.7 & $28.6 \pm 4.4$ & 21.9/31 & $40.2 \pm 0.8$ & 2.2 & $8.0 \pm 1.1$ & 1.3\\
   & $f_{h}$ & $81.1 \pm 3.0$ &  5.4 & $ 8.9 \pm 2.0$ & 3.2 & -              & -       & $90.4 \pm 2.1$ & 4.9 & $3.6 \pm 0.7$ & 1.6\\
06 & $f_{1}$ & $35.7 \pm 1.0$ &  6.7 & $13.4 \pm 3.3$ & 2.0 & $24.3 \pm 5.2$ & 39.3/29 & $35.6 \pm 0.7$ & 7.8 & $4.5 \pm 0.8$ & 2.5\\
   & $f_{2}$ & $46.1 \pm 0.9$ & 17.1 & $12.3 \pm 3.1$ & 2.3 & -              & -       & $45.0 \pm 1.5$ & 5.1 & $4.3 \pm 0.8$ & 2.8\\
   & $f_{h}$ & $89.0 \pm 2.8$ & 11.8 & $ 7.5 \pm 2.2$ & 1.1 & -              & -       & $83.6 \pm 2.0$ & 4.0 & $3.4 \pm 0.6$ & 1.6\\
07 & $f_{1}$ & $34.0 \pm 1.2$ &  5.8 & $14.2 \pm 3.2$ & 3.6 & $26.7 \pm 5.5$ & 29.4/31 & $33.7 \pm 0.6$ & 4.7 & $4.9 \pm 1.0$ & 1.5\\
   & $f_{2}$ & $45.1 \pm 1.6$ &  5.0 & $15.3 \pm 3.1$ & 2.2 & -              & -       & $44.8 \pm 0.8$ & 5.5 & $5.5 \pm 1.1$ & 1.6\\
   & $f_{h}$ & $77.9 \pm 2.2$ &  5.9 & $ 9.0 \pm 2.1$ & 4.8 & -              & -       & $83.5 \pm 1.7$ & 3.3 & $3.7 \pm 0.7$ & 1.3\\
08 & $f_{1}$ & $33.1 \pm 1.2$ &  4.5 & $13.9 \pm 3.1$ & 1.9 & $24.8 \pm 3.9$ & 26.3/22 & $33.9 \pm 1.6$ & 3.9 & $3.6 \pm 0.6$ & 2.0\\
   & $f_{2}$ & $43.6 \pm 1.1$ &  5.9 & $16.1 \pm 3.6$ & 2.9 & -              & -       & $43.3 \pm 1.2$ & 6.2 & $4.4 \pm 0.8$ & 1.8\\
   & $f_{h}$ & $84.5 \pm 2.1$ &  5.3 & $ 8.4 \pm 1.9$ & 1.6 & -              & -       & $86.5 \pm 5.7$ & 1.4 & $3.2 \pm 0.5$ & 1.3\\
09 & $f_{1}$ & $38.0 \pm 1.3$ &  3.8 & $17.8 \pm 3.1$ & 1.9 & $23.4 \pm 4.3$ & 28.3/31 & $37.6 \pm 0.8$ & 4.0 & $5.6 \pm 0.9$ & 1.4\\
   & $f_{h}$ & $64.9 \pm 4.2$ &  5.5 & $ 7.7 \pm 1.9$ & 1.1 & -              & -       & $70.0 \pm 2.5$ & 1.6 & $2.9 \pm 0.5$ & 1.5\\ 
10 & $f_{1}$ & $38.8 \pm 1.0$ &  9.9 & $15.8 \pm 3.0$ & 2.8 & $23.8 \pm 3.5$ & 64.1/34 & $34.0 \pm 1.4$ & 2.3 & $4.6 \pm 0.6$ & 1.6\\
   & $f_{h}$ & $73.4 \pm 2.7$ &  4.7 & $ 8.7 \pm 1.8$ & 2.2 & -              & -       & $70.6 \pm 1.2$ & 1.8 & $2.1 \pm 0.5$ & 3.2\\
11 & $f_{1}$ & $28.1 \pm 1.0$ &  3.1 & $14.4 \pm 3.6$ & 1.7 & $28.4 \pm 5.6$ & 12.5/11 & $27.3 \pm 1.1$ & 2.8 & $4.2 \pm 0.8$ & 1.7\\
   & $f_{2}$ & $40.5 \pm 0.6$ & 15.0 & $11.6 \pm 3.3$ & 1.7 & -              & -       & $38.1 \pm 0.9$ &12.6 & $3.9 \pm 0.9$ & 5.1\\
   & $f_{h}$ & $66.0 \pm 4.7$ &  2.1 & $14.7 \pm 3.3$ & 1.7 & -              & -       & $65.5 \pm 1.8$ & 3.3 & $2.9 \pm 0.5$ & 1.4\\   
12 & $f_{1}$ & $25.6 \pm 1.7$ &  3.3 & $15.8 \pm 3.2$ & 1.6 & $19.4 \pm 3.6$ & 14.8/18 & $27.5 \pm 1.4$ &11.5 & $2.3 \pm 0.5$ & 2.2\\
   & $f_{h}$ & $87.2 \pm 2.1$ & 19.8 & $ 7.6 \pm 1.7$ & 6.7 & -              & -       & $87.8 \pm 1.7$ &16.6 & $1.5 \pm 0.3$ & 1.9\\
13 & $f_{1}$ & $27.3 \pm 0.5$ &  6.2 & $10.8 \pm 2.4$ & 1.9 & $17.1 \pm 3.2$ & 29.6/34 & $26.4 \pm 1.0$ & 3.0 & $3.9 \pm 0.6$ & 1.5\\
   & $f_{h}$ & $70.9 \pm 3.6$ &  7.5 & $ 3.5 \pm 0.8$ & 6.7 & -              & -       & $77.0 \pm 1.2$ &11.4 & $1.5 \pm 0.3$ & 2.4\\ 
14 & $f_{1}$ & $27.7 \pm 0.4$ &  8.1 & $14.0 \pm 2.9$ & 2.5 & $18.4 \pm 3.4$ & 41.8/35 & $27.1 \pm 0.7$ & 6.3 & $3.6 \pm 0.7$ & 2.8\\
   & $f_{h}$ & $84.2 \pm 2.6$ & 17.2 & $ 4.8 \pm 1.2$ & -   & -              & -       & $83.3 \pm 6.7$ &11.4 & $1.5 \pm 0.3$ & 2.4\\
15 & $f_{1}$ & $28.2 \pm 0.5$ &  6.1 & $12.7 \pm 2.6$ & 1.6 & $16.5 \pm 3.1$ & 51.8/34 & $27.6 \pm 0.6$ & 7.1 & $2.6 \pm 0.6$ & 2.3\\
   & $f_{h}$ & $55.3 \pm 2.8$ &  4.8 & $ 4.1 \pm 1.0$ & 4.0 & -              & -       & $50.5 \pm 3.2$ & 2.3 & $2.6 \pm 0.5$ & 1.8\\
16 & $f_{1}$ & $30.3 \pm 0.6$ &  6.8 & $12.1 \pm 2.7$ & 2.3 & $16.7 \pm 3.2$ & 68.4/37 & $28.7 \pm 1.3$ & 2.4 & $3.6 \pm 0.6$ & 1.6\\
   & $f_{h}$ & $66.1 \pm 0.5$ &  4.6 & $ 4.4 \pm 1.5$ & 2.6 & -              & -       & $67.9 \pm 2.2$ & 4.5 & $1.6 \pm 0.4$ & 1.9\\
17 & $f_{1}$ & $26.0 \pm 1.0$ &  4.8 & $10.3 \pm 2.6$ & 1.6 & $18.7 \pm 3.6$ & 15.9/18 & $23.6 \pm 1.0$ & 8.3 & $2.3 \pm 0.5$ & 6.8\\
   & $f_{2}$ & $33.7 \pm 1.7$ & 11.5 & $ 9.5 \pm 2.4$ & 2.4 & -              & -       & $31.5 \pm 0.8$ & 5.5 & $3.0 \pm 0.7$ & 2.1\\
   & $f_{h}$ & $63.9 \pm 3.7$ &  5.2 & $ 6.2 \pm 1.4$ & 1.2 & -              & -       & $66.6 \pm 2.5$ & 4.5 & $1.7 \pm 0.4$ & 1.9\\
18 & $f_{1}$ & $27.6 \pm 0.5$ &  7.9 & $13.6 \pm 2.8$ & 1.9 & $18.2 \pm 3.3$ & 35.4/26 & $31.6 \pm 0.5$ &12.2 & $3.2 \pm 0.6$ & 1.4\\
   & $f_{h}$ & $57.5 \pm 1.8$ &  6.4 & $ 5.8 \pm 1.3$ & 2.3 & -              & -       & $56.0 \pm 0.9$ &26.2 & $1.3 \pm 0.3$ & 3.9\\
19 & $f_{1}$ & $28.2 \pm 0.7$ &  6.1 & $12.7 \pm 2.6$ & 2.9 & $18.2 \pm 3.0$ & 46.7/37 & $28.6 \pm 3.3$ &12.5 & $3.2 \pm 0.7$ & 1.7\\
   & $f_{h}$ & $54.0 \pm 1.8$ &  8.9 & $ 5.8 \pm 1.5$ & 1.1 & -              & -       & $59.1 \pm 2.2$ & 4.6 & $2.2 \pm 0.5$ & 1.7\\ 
		 \hline
	\end{tabular}
\end{table*}

\begin{figure}
	{
		\includegraphics[scale=0.32,angle=-90]{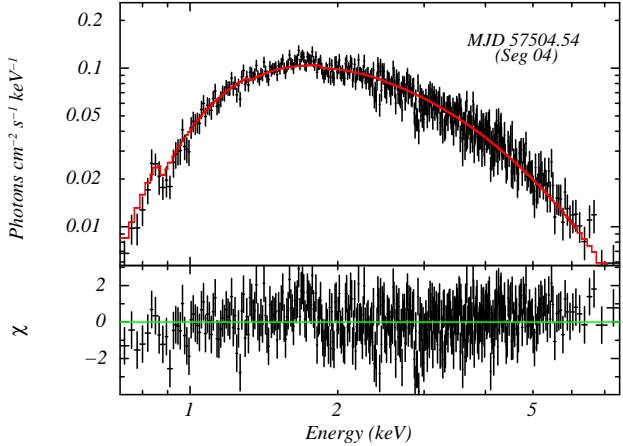}		
	}
	\caption{\textit{SXT} spectrum of IGR J17091--3624 for the energy range 0.7--7 keV for the Segment 4. The spectrum was best fitted with Model 1 (\texttt{TBabs(ezdiskbb)}).}
	\label{fig:7}
\end{figure}

\begin{table*}
	\centering
	\caption{Best fit model parameters and estimated values of \textit{SXT} spectra (0.7--7 keV) of IGR J17091--3624 with the Model 1: \texttt{TBabs(ezdiskbb)} in comparison with fitting the \textit{SXT} spectrum of GRS 1915+105 in the same energy range. $\hbox{Norm} - (1/f^4) (R_{in}/D)^2 \cos i$, where $R_{in}$ is the disk inner radius in km, $D$ is the source distance in units of 10 kpc, $i$ is the inclination, and $f$ is the colour to effective temperature ratio.}
	\label{tab:3}
	\begin{tabular}{cccccc} 
		\hline
	\multicolumn{6}{c}{\textit{IGR J17091--3624}}\\
	\hline
	Seg. & $N_{H}$ & T$_{max}$ & $Norm$ & $\chi_{red}^{2}$ & Flux (2-10 keV)\\
	(No) & $10^{22}$ \textit{atoms/cm}$^{2}$ & (keV) & & ($\chi^{2}$/\textit{dof}) & $10^{-9}$ergs/cm$^{2}$/s\\
	\hline
	1 & $0.85\pm 0.02$ & $1.28\pm 0.02$ & $6.57\pm 0.5$ & 445/384 & $1.24\pm0.01$\\
	2 & $0.83\pm 0.03$ & $1.29\pm 0.02$ & $6.13\pm 0.5$ & 419/410 & $1.23\pm0.01$\\
	3 & $0.86\pm 0.03$ & $1.23\pm 0.02$ & $7.27\pm 0.5$ & 466/410 & $1.77\pm0.01$\\
	4 & $0.83\pm 0.02$ & $1.27\pm 0.02$ & $6.63\pm 0.5$ & 460/409 & $1.77\pm0.01$\\
	5 & $0.80\pm 0.03$ & $1.26\pm 0.02$ & $6.58\pm 0.5$ & 515/410 & $1.17\pm0.01$\\
	6 & $0.86\pm 0.03$ & $1.25\pm 0.03$ & $6.85\pm 0.6$ & 353/353 & $1.16\pm0.01$\\
	7 & $0.80\pm 0.03$ & $1.28\pm 0.02$ & $6.13\pm 0.5$ & 441/411 & $1.18\pm0.01$\\
	8 & $0.81\pm 0.02$ & $1.26\pm 0.02$ & $6.64\pm 0.5$ & 542/415 & $1.17\pm0.01$\\
	9 & $0.86\pm 0.03$ & $1.23\pm 0.02$ & $7.20\pm 0.6$ & 396/355 & $1.14\pm0.01$\\
   10 & $0.83\pm 0.03$ & $1.29\pm 0.03$ & $5.87\pm 0.5$ & 369/369 & $1.17\pm0.01$\\
   11 & $0.81\pm 0.04$ & $1.26\pm 0.03$ & $6.11\pm 0.6$ & 348/326 & $1.10\pm0.01$\\
   12 & $0.80\pm 0.03$ & $1.31\pm 0.03$ & $5.57\pm 0.6$ & 257/301 & $1.18\pm0.01$\\
   13 & $0.81\pm 0.03$ & $1.26\pm 0.03$ & $6.32\pm 0.6$ & 372/342 & $1.13\pm0.01$\\
   14 & $0.84\pm 0.03$ & $1.26\pm 0.03$ & $6.67\pm 0.6$ & 393/355 & $1.18\pm0.01$\\
   15 & $0.84\pm 0.03$ & $1.25\pm 0.02$ & $7.00\pm 0.5$ & 521/411 & $1.17\pm0.01$\\
   16 & $0.85\pm 0.03$ & $1.24\pm 0.02$ & $7.03\pm 0.5$ & 449/411 & $1.17\pm0.01$\\
   17 & $0.82\pm 0.03$ & $1.27\pm 0.02$ & $6.31\pm 0.5$ & 453/405 & $1.19\pm0.01$\\
   18 & $0.80\pm 0.02$ & $1.28\pm 0.02$ & $6.14\pm 0.4$ & 397/411 & $1.19\pm0.01$\\
   19 & $0.85\pm 0.02$ & $1.21\pm 0.02$ & $7.56\pm 0.5$ & 485/401 & $1.13\pm0.01$\\
	\hline
   \multicolumn{5}{c}{\textit{GRS 1915+105}}\\
   \hline
    1 & $5.1\pm 0.1$ & $1.28\pm 0.02$ & $40.0\pm 3.0$ & 507/418 & $7.87\pm0.03$\\
   \hline
 	\end{tabular}
\end{table*}  

\begin{table*}
	\setlength{\tabcolsep}{5pt}
	\centering
\caption{Best fit broadband model parameters and estimated values of both \textit{SXT \& LAXPC} spectra (0.7--23 keV) of IGR J17091--3624 and GRS 1915+105 with the Model 2a: \texttt{TBabs(ezdiskbb+gaussian+nthcomp)} for IGR J17091--3624 and Model 2b: \texttt{TBabs(ezdiskbb+diskline+nthcomp)} for GRS 1915+105. The $N_{H}$ values are expressed in units of $10^{22}$ atoms cm$^{-2}$, whereas the $Flux$ in the energy range 2--10 keV is expressed in units of \textit{$10^{-9}$ergs/cm$^{2}$/s}.}
	\label{tab:4}.
	\begin{tabular}{cccccc} 
	\hline
	\multicolumn{5}{c}{\textit{IGR J17091--3624}} & \textit{GRS 1915+105}\\
	\hline
	& \multicolumn{4}{c}{Model 2a: \texttt{TBabs(ezdiskbb+gaussian+nthcomp)}} &  Model 2b: \texttt{TBabs(ezdiskbb+diskline+nthcomp)}\\
	\hline
	Parameters     &Seg. 3 (SPQ)   &Seg. 7 (DPQ)   &Seg. 11 (DPQ)  &Seg. 16 (SPQ)  &Orbit No. 8118\\
	\hline
	$N_{H}$        & $0.86^{f}$    & $0.80^{f}$    & $0.81^{f}$    & $0.85^{f}$    & $5.4\pm0.2$\\
	$T_{max}$ (keV)& $0.78\pm0.02$ & $0.84\pm0.03$ & $0.72\pm0.03$ & $0.63\pm0.03$ & $0.95\pm0.01$\\
	$Norm$         & $7.27^{f}$    & $6.13^{f}$    & $6.11^{f}$    & $7.03^{f}$    & $86\pm9$    \\
	$\Gamma$       & $2.37\pm0.02$ & $2.40\pm0.03$ & $2.23\pm0.02$ & $2.13\pm0.02$ & $2.16\pm0.02$\\
	$kT_e$ (keV)   & $4.02\pm0.11$ & $3.80\pm0.08$ & $3.66\pm0.07$ & $3.51\pm0.06$ & $41.0\pm12.0$ \\
	$Norm$         & $0.79\pm0.03$ & $0.79\pm0.03$ & $0.90\pm0.04$ & $1.08\pm0.03$ & $0.84\pm0.04$\\
	$\chi^{2}_{red}$ ($\chi^{2}$/\textit{dof}) & 560/438 & 501/433 & 439/356 & 545/437 & 560/473  \\
	$Flux$         & $3.60\pm0.003$& $3.77\pm0.003$& $4.02\pm0.003$& $4.64\pm0.003$& $8.77\pm0.04$ \\
	\hline
	\footnotesize{$f$ - \textit{The values are frozen.}}
	\end{tabular}
\end{table*}

\begin{table*}
	\centering
	\caption{Fit parameters of the PDS obtained from \textit{SXT} and \textit{LAXPC} data with multiple Lorentzian for GRS 1915+105.}
	\label{tab:time_var}
	\begin{tabular}{ccccccccc} 
		\hline
		\multicolumn{9}{c}{\textit{SXT}}\\
		\hline
		\multicolumn{6}{c}{`Heartbeat' QPO} & Total\\
		&&Frequency (mHz) & Q$_{factor} $ & Significance ($\sigma$) & rms (\%) & rms (\%)\\
		\hline
		&& 6.7$^{+0.9}_{-0.8}$ ($f_{1}$) & 5.6 & 2.4 & 9.2$\pm$1.5 & 13.1$\pm$1.8\\
		&& 14.0$^{+0.1}_{-0.1}$ ($f_{h}$) & 2.1 & 1.9 & 6.5$\pm$1.7 & & \\
		\hline
		\hline
		\multicolumn{9}{c}{\textit{LAXPC}}\\
		\hline
		\multicolumn{4}{c}{`Heartbeat' QPO} & \multicolumn{4}{c}{QPO} & Total \\
		Frequency (mHz) & Q$_{factor} $ & Significance ($\sigma$) & rms (\%) & Frequency (Hz) & Q$_{factor} $ & Significance ($\sigma$) & rms (\%) & rms (\%)\\
		\hline
		6.2$^{+1.0}_{-1.1}$ ($f_{1}$) & 7.9 & 6.8 & 13.8$\pm$1.2 & 5.2$^{+0.09}_{-0.06}$ & 4.2 & 10.5 & 8.7$\pm$0.7 & 23.4 $\pm$ 1.5\\	
		14.0$^{+0.1}_{-0.1}$ ($f_{h}$) & 4.7 & 2.4 & 6.1$\pm$1.6 & & \\
		\hline
	\end{tabular}
\end{table*}

\begin{table*}
	\centering
\caption{Best fit broadband model parameters and estimated values of both \textit{SXT \& LAXPC} phase-resolved spectra ($ 0.7-23 $ keV) with the Model 2a : \texttt{TBabs(ezdiskbb+nthcomp+gaussian)} for IGR J17091--3624 and Model 2b : \texttt{TBabs(ezdiskbb+diskline+nthcomp)} for GRS 1915+105. The $N_{H}$ value are expressed in units of $10^{22}$ atoms cm$^{-2}$, whereas the $Flux$ in the energy range 2--10 keV is expressed in units of $10^{-9}$ erg cm$^{-2}$ s$^{-1}$.}
	\label{tab:6}.
	\begin{tabular}{cccccccc} 
		\hline
		 & \multicolumn{4}{c}{\textit{IGR J17091--3624}} & \multicolumn{3}{c}{\textit{GRS 1915+105}}\\
		& \multicolumn{2}{c}{SPQ} & \multicolumn{2}{c}{DPQ} & \multicolumn{3}{c}{}\\
		\hline
		Parameters       & Seg. 3 (I) & Seg. 3 (II) & Seg. 7 (I) & Seg. 7 (II)  & (P$_{a}$)    & (P$_{b}$)     & (P$_{c}$)\\
		\hline                                
		$N_{H}$          & $0.86^{f}$    & $0.90^{f}$    & $0.80^{f}$    & $0.79^{f}$    & $5.0\pm0.4$  & $4.9\pm0.2$   & $5.4\pm0.4$\\
		$T_{max}$ (keV)  & $0.68\pm0.03$ & $1.19\pm0.08$ & $0.72\pm0.03$ & $1.22\pm0.09$ & $1.20\pm0.1$ & $1.11\pm0.04$ & $1.2\pm0.1$\\
        $Norm$           & $8.36^{f}$    & $6.70^{f}$    & $6.60^{f}$    & $4.16^{f}$    & $24.0\pm9.0$ & $39.0\pm7.0$  & $40.0\pm16.0$\\
		$\Gamma$         & $2.34\pm0.02$ & $2.52\pm0.06$ & $2.31\pm0.02$ & $2.67\pm0.08$ & $2.06\pm0.06$& $2.06\pm0.02$ & $2.1\pm0.2$\\
		$kT_e$ (keV)     & $4.06\pm0.11$ & $4.18\pm0.26$ & $3.70\pm0.08$ & $4.21\pm0.30$ & $41.0^{f}$   & $30.0\pm6.0$  & $12.0\pm7.0$ \\
		$Norm$           & $0.85\pm0.04$ & $0.56\pm0.10$ & $0.82\pm0.03$ & $0.65\pm0.08$ & $0.41\pm0.09$& $0.56\pm0.06$ & $0.6\pm0.3$\\
		$\chi^{2}_{red}$ & $ 510/380 $   & 224/233       & 501/395       & 266/244       & 107/99       & 424/394       & 93/106       \\
		$Flux$           & $3.33\pm0.003$& $4.54\pm0.004$& $3.47\pm0.003$& $4.67\pm0.004$& $6.08\pm0.07$& $7.16\pm0.05$ & $10.5\pm0.1$\\
		\hline
		\footnotesize{$f$ - \textit{The values are frozen.}}		
	\end{tabular}
\end{table*}

\subsection{Phase-Resolved Spectroscopy}
 \label{sec:phas}
 In this section, we present the phase-resolved spectroscopic properties of IGR J17091--3624 \citep[see][]{Rao2012} with those of single peak QPO and double peak QPO in the `heartbeat' State.
 
Phase-resolved spectroscopy (PRS) for IGR J17091-3624 light curve was carried out for the segments 3 and 7 into two phases: Plateau region (Phase I) and Peak region (Phase II), as the intermediate region was not observed due to short and random nature of the pulse profile. Considering a 1 s binned \textit{LAXPC} light curve, these two regions were bifurcated based on the count rate threshold as shown in the upper panel of the Fig. \ref{fig:prs}. Phase I light curve events, with the threshold count rate less than 750 counts s$^{-1}$, are marked with the pink band, whereas the Phase II events, having the threshold count rate greater than 800 counts s$^{-1}$ are marked with the violet band. A small region between burst and plateau phase was filtered out to ignore small stray peaks in between. To produce these phase resolved light curves, we generated individual GTI's using a FORTRAN code. These GTI's are then given as input to the \textit{LAXPC} pipeline to regenerate the Level 2 spectrum as described in Sect.~\ref{obs}.

We model the PRS using Model 2a applied to the time averaged spectrum. The phases and the variation of parameters of segment 3 are shown in Fig. \ref{fig:prs}. We find very little variation in parameters across the two phases. There is a slight increase in photon index and electron plasma temperature for Phase II along with almost double rise in the disc temperature in both single and double peak QPO's, indicates that the spectrum becomes softer in Phase II. The phase parameter variations of segments 3 and 7 of broadband results are presented in Table \ref{tab:6}. The estimated flux for 2--10 keV was in the range of 3.33--4.67~$\times  10^{-9}$ ergs cm$^{-2}$ s$^{-1}$. A decrease in the energy of \texttt{gaussian} line is observed during the Phase II in both single and double peak QPO segments. 

\begin{figure}
	{
		\includegraphics[scale=0.32,angle=-90]{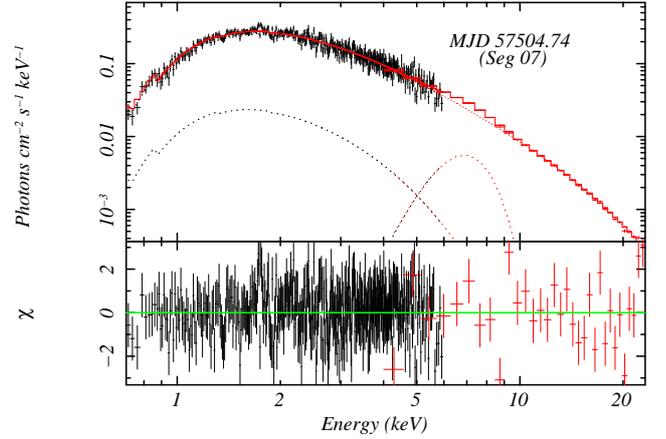}	
	}
	\caption{Broadband \textit{SXT} and \textit{LAXPC} spectra of IGR J17091--3624 for the energy range 0.7--23 keV for the Segment 7. The spectrum was best fitted with Model 2a (\texttt{TBabs(ezdiskbb+gaussian+nthcomp)}) and residuals are plotted in the bottom panel in units of $\sigma$ (see text for details). The dashed lines are the contributions of the model components.}
	\label{fig:lxpspc}
\end{figure}

\begin{figure*}
	\includegraphics[scale=0.5]{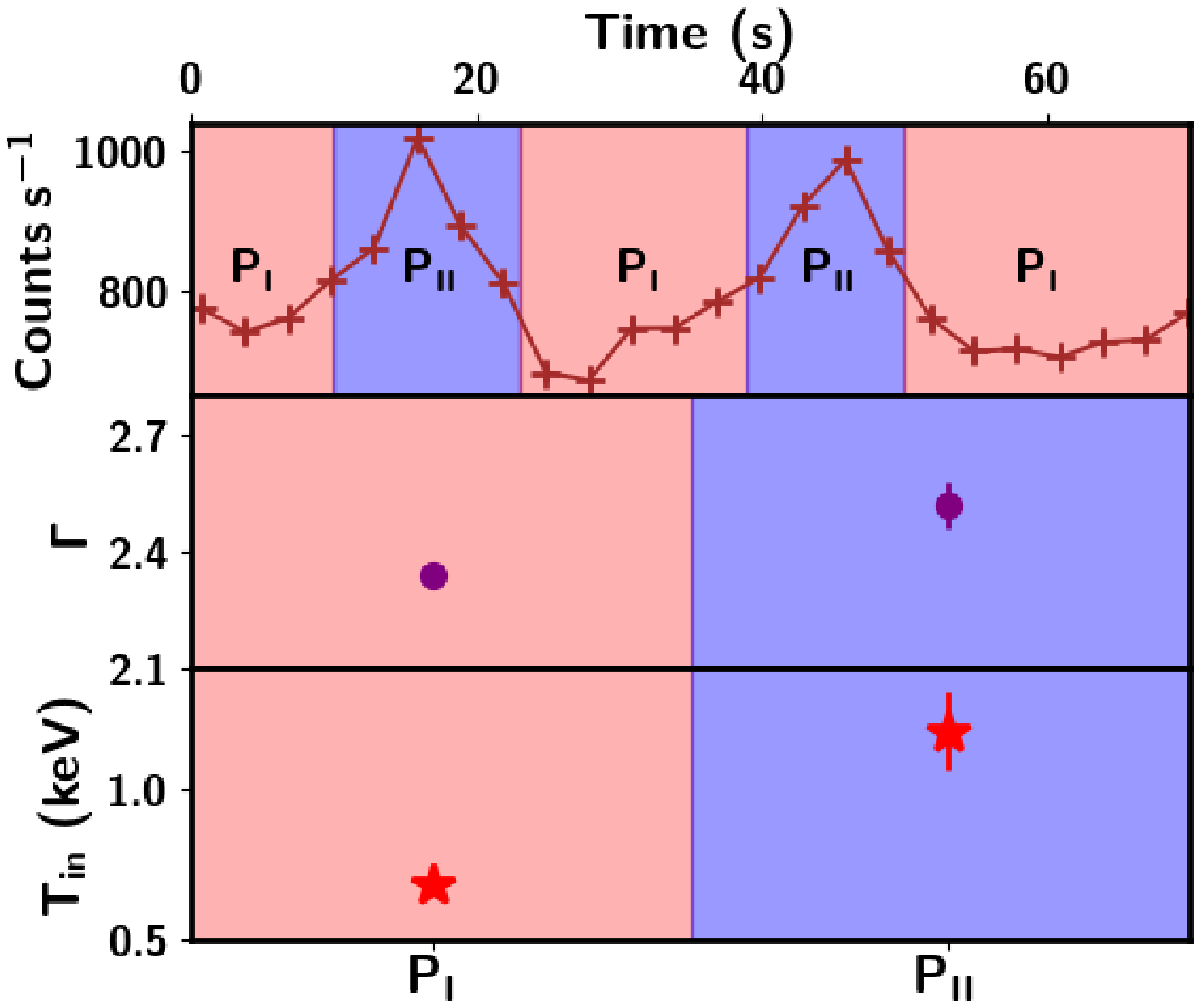}
	\hspace{1cm}
	\includegraphics[scale=0.5]{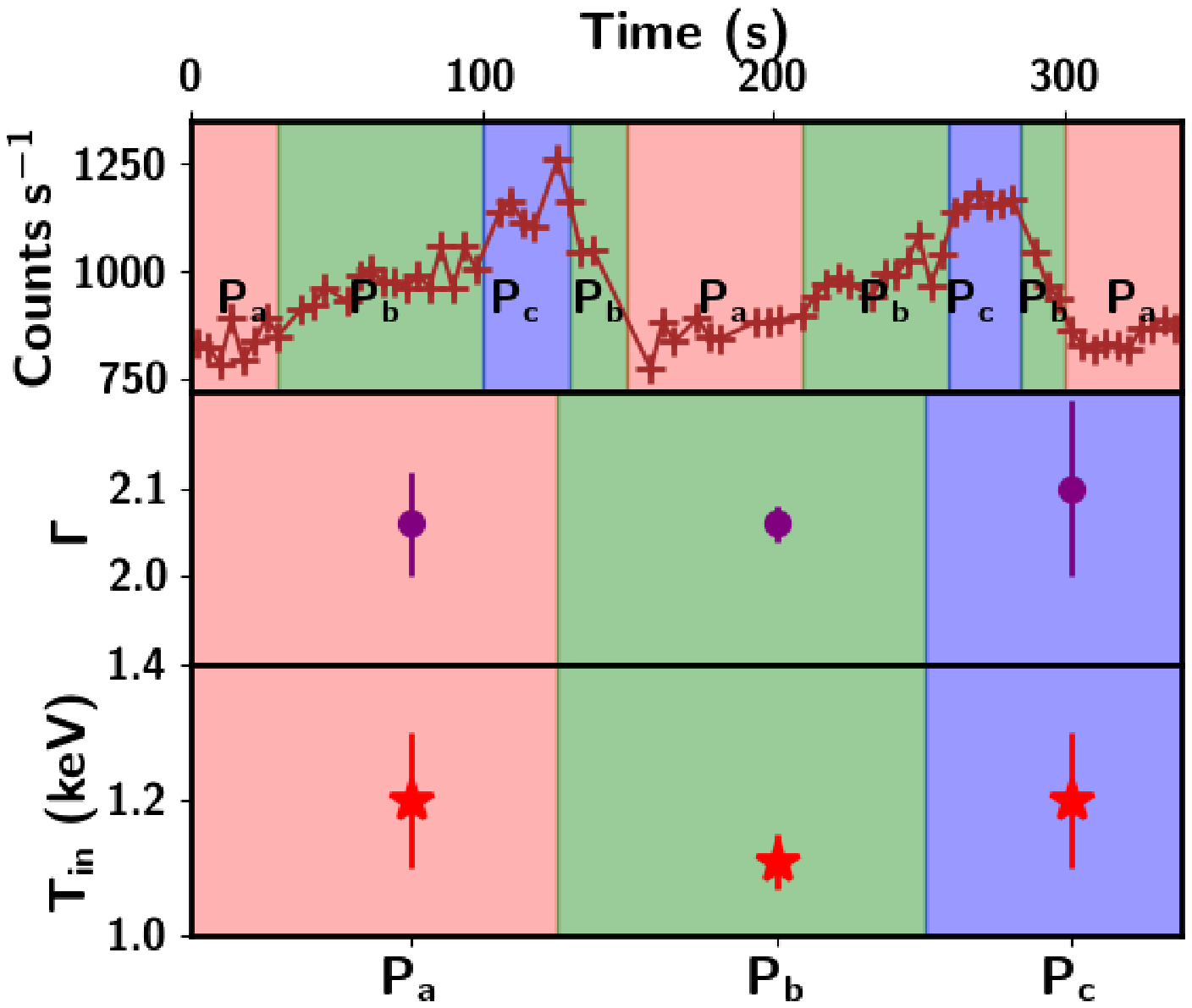}
	\caption{The left panel represent the \textit{LAXPC20} phase resolved light curve of IGR J17091--3624 for the Segment 3 and right panel represent the GRS 1915+105. The lower panels display the average broadband  spectral parameters of both \textit{SXT} and \textit{LAXPC} such as photon index (purple circles) and disc temperature (red stars) of the PRS for Segment 3 of the IGR J17091--3624 (left) and GRS 1915+105 (right).}
	\label{fig:prs}
\end{figure*}

\begin{figure}
	\includegraphics[scale=0.35]{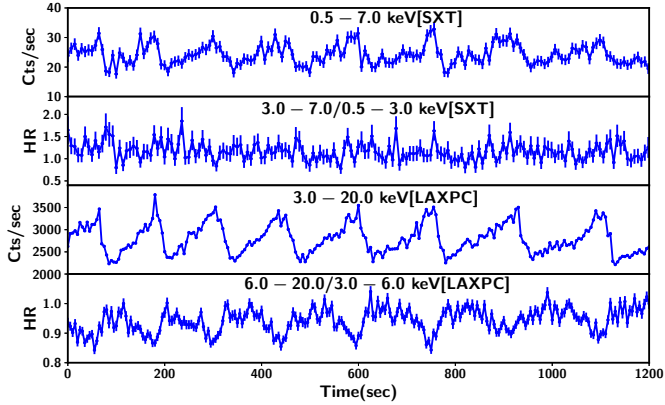}
	\caption{The light curves of GRS 1915+105 observed with \textit{SXT} and \textit{LAXPC} are shown in panel 1 and 3, respectively. The \textit{SXT} and \textit{LAXPC} light curves are binned by 7.134 s and 5 s for better representation of the `HS' state. The hardness ratios are plotted in panel 2 (\textit{SXT}) and 4 (\textit{LAXPC}).}
	\label{fig:8}
\end{figure}

\begin{figure}
 \centering
	\includegraphics[scale=0.5]{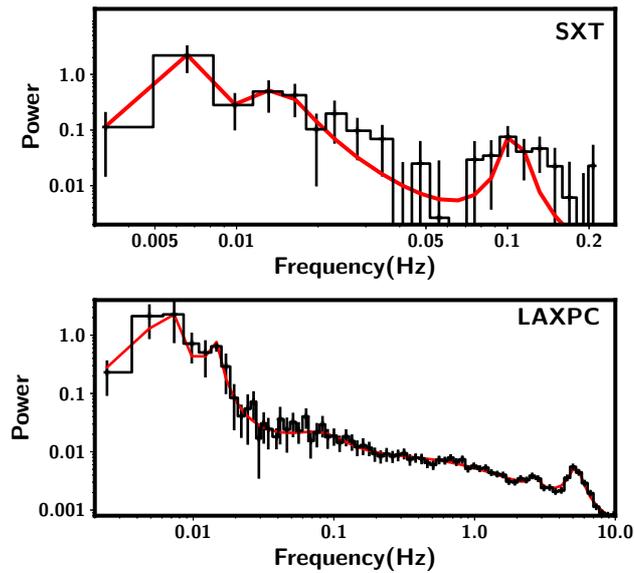}
	\caption{PDS from \textit{SXT} and \textit{LAXPC} data of GRS 1915+105 are plotted in the top and bottom panels respectively. A QPO at 5 Hz and its sub-harmonic at $\sim 2.5$ Hz are clearly visible in the \textit{LAXPC} PDS along with the `heartbeat' QPO at 7 mHz and its harmonic at 14 mHz.}
	\label{fig:9}
\end{figure}

\begin{figure}
	\includegraphics[angle=-90,width=\columnwidth]{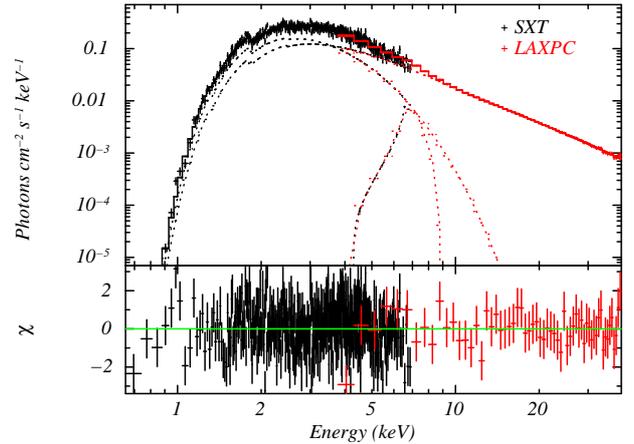}
	\caption{Broadband \textit{SXT} and \textit{LAXPC} spectral data of GRS 1915+105 is fitted with the model \texttt{Tbabs(ezdiskbb+diskline+nthcomp)}. Residuals are plotted in the bottom panel in units of $\sigma$. The dashed lines are the contributions of the model components.}
	\label{fig:10}
\end{figure}

\subsection{Comparison with GRS 1915+105}
 \label{sec:comp}

In this section, we compare the timing, spectral and PRS properties of IGR J17091--3624 with those of GRS 1915+105 in the `heartbeat' State.

\subsubsection{Timing Properties of GRS 1915+105}
Till date, only one observation with \textit{AstroSat} of GRS 1915+105 is reported to be in the `heartbeat' state \citep{Rawat2019}. The \textit{SXT} light curve in the 0.5--7 keV band is shown in top panel of Fig.~\ref{fig:8}. Hardness ratio as defined in Sect.~\ref{sec:hid} is shown in next panel. `Heartbeat' variability is also seen in the \textit{SXT} light curve (top panel of Fig.~\ref{fig:8}), which is more clear and prominent in the \textit{LAXPC} light curve (third panel of Fig.~\ref{fig:8}). The hardness ratio clearly shows an anti-correlation with the flux as is clear from the two lower panels of Fig.~\ref{fig:8}. The PDS of SXT light curve is generated with a time resolution of 2.378 s in the frequency range 3 mHz to 0.25 Hz.  
The PDS for \textit{LAXPC} is obtained in the 3--20 keV band for a time resolution of 50 ms in the 2 mHz to 10 Hz frequency range. The model fitted PDS are shown in Fig.~\ref{fig:9} and the fitted parameters along with estimated rms are summarized in Table \ref{tab:time_var}. 

\textit{SXT} PDS of GRS 1915+105 is shown in the top panel of Fig.~\ref{fig:9} which shows a `heartbeat' QPO at $\sim 7$ mHz along with an associated harmonic at $\sim 14$ mHz. The `heartbeat' QPO is prominent in the \textit{SXT} PDS with an rms of 9.2\% along with a harmonic contributing to 6\% of the total rms of 13\%. These features are also seen in the \textit{LAXPC} PDS. The QPO at $\sim 5 $ Hz is seen in the \textit{LAXPC} PDS along with its sub-harmonic at $\sim 2.5$ Hz while these features are either absent or not discernible in the PDS of IGR J17091--3624 due to the LAXPC PDS being noise dominated beyond 1 Hz. The heartbeat QPO at $ \sim7 $ mHz contributes to 13.8\% of the total rms of 23.4\% in the 2 mHz to 10 Hz frequency range. Light curves of the two sources IGR J17091--3624 and GRS 1915+105, differ in periodicity and structure of the burst profile as seen in Figs.~\ref{fig:4},~\ref{fig:5} and \ref{fig:8}. The double peak feature observed in IGR J17091--3624 is not seen in GRS 1915+105. 

\subsubsection{Spectral Properties of GRS 1915+105}

Following the fitting procedure to the spectra of IGR J17091-3624, we first fit the \textit{SXT} spectrum of GRS 1915+105 independently from 0.7--7 keV. This could be modelled using Model 1 with a disc temperature of 1.28 keV and norm as $\sim 40$ (see Table \ref{tab:3}). Although the disc temperatures are similar, the norm for GRS 1915+105 is obtained as 40, which is $\sim$ 6 times larger than the one for IGR J17091--3624, implying a larger inner disc radius. The simultaneous \textit{SXT} and \textit{LAXPC} broadband spectrum was modelled in the energy range of 0.7--40 keV \textbf{with the Model 2a that gave a $\chi_{red}^2 = 1.19~(562/472)$}. Replacing the \texttt{gaussian} with \texttt{diskline} and adding a \texttt{smedge} resulted in slightly better fit with $\chi_{red}^{2} = 1.18~(560/473)$. The centroid energy was obtained at $6.8 \pm 0.2$ keV. However, better constraints on other parameters could not be obtained and were frozen to default values. The fitted parameters are given in Table \ref{tab:4}. Assuming the disc to be the source of the input seed photons, the $kT_{in}$ parameter of \texttt{nthcomp} model was tied to the disc temperature of \texttt{ezdiskbb} which was obtained close to $\sim 1$ keV. Although the photon indices for both sources are comparable (2.1--2.4 for IGR J17091--3624 and 2.1 for GRS 1915+105), the disc temperature of IGR J17091-3624 is \textbf{between 66\% and 88\%} of that of GRS 1915+105 and the electron plasma temperature is an order of magnitude lower (see Table \ref{tab:4}).

\subsubsection{Phase-Resolved Spectroscopy of GRS 1915+105}

PRS of GRS 1915+105 was made for comparison with IGR J17091-3624. Light curve of GRS 1915+105 is relatively more structured compared to the one from IGR J17091-3624 as seen from Fig. \ref{fig:4}, \ref{fig:5} and \ref{fig:prs}. Hence, the light curve could be divided into three phases instead of the two considered for IGR J17091-3624. Flux thresholds are defined based on the structure of the light curve wherein the events with count rate less than 900 counts s$^{-1}$ belong to Phase P$_{a}$, those lying above 1100 counts s$^{-1}$ lie in Phase P$_{c}$ while the intermediate events belong to Phase P$_{b}$.

We model the PRS using Model 2b applied to the time averaged spectrum. The phases and the variation of parameters are shown in Fig. \ref{fig:prs} (bottom-right panel). We find very little variation in parameters across the three phases. There is a slight increase in photon index in Phase P$_{c}$, suggesting that the spectrum becomes softer in this phase, although the disc temperature remains relatively constant. The change in parameters is shown in Table \ref{tab:6}. The electron plasma temperature remains unconstrained for Phase P$_{a}$ and is hence frozen to the value obtained from the time-averaged spectrum. A decrease in $kT_e$ is then observed from Phase P$_{b}$ to Phase P$_{c}$. \textbf{A decrease in the equivalent width of the emission line from 220 to 187 eV is also seen.}

\section{Discussions and Conclusion}
\label{dis}
In this paper, we carried out extensive temporal and spectral analysis of the black hole source IGR J17091--3624 during its 2016 outburst observed with \textit{SXT} and \textit{LAXPC} onboard \textit{AstroSat}. IGR J17091--3624 was the first BHB to exhibit some of the variabilities which were earlier unique only to GRS 1915+105 prompting extensive publications on comparisons of this source with GRS 1915+105 and describing the accretion process based on phenomenological and theoretical models \citep{Anuj2001, Bell2005, Dunn2010, Alta2011, Capi2012, Nandi2012, Zhang2014, Radhi2014, Pahari2014} including a two component advective flow model \citep{Iyer2015,Radhi2018}. During the \textit{AstroSat} observation, we found IGR J17091--3624 to be in the `heartbeat' class similar to the one found in GRS 1915+105. Therefore, we made a comparative study of such variability with GRS 1915+105 observation from the \textit{AstroSat} data in March 2017 \citep{Rawat2019}.  

IGR J17091--3624 was in steady soft-intermediate state during the \textit{AstroSat} observation, as seen from the Fig.~\ref{fig:3} \citep[see][for details]{Radhi2018}. The source has shown repetitive and high variability with hardness ratio varying from 0.2 to 0.7 (except in a few cases, where HR $\sim1$) and with count rate varying from $\sim5$ to 35 c/s as shown in Figs.~\ref{fig:3}, \ref{fig:4} and \ref{fig:5} similar to the variability seen in GRS 1915+105 (see Fig.~\ref{fig:8}). However, the variability timescales in IGR J17091--3624 lies within 20--40 s, which is faster than the timescale ($ \sim150 $ s) of GRS 1915+105. The hardness ratio for both sources shows a strong anti-correlation with the flux during the `heartbeat' class as was previously observed by \cite{Pahari2014}. Also, the average values of hardness ratio observed in \textit{SXT} is higher in GRS 1915+105 than that in IGR J17091--3621 signalling a higher contribution in the latter source from the Keplerian disc.

For the first time, we find two different types of `heartbeat' QPOs in the PDS for the source IGR J17091--3624 --- single and double peak spanning in the frequency range between 25--48 mHz (as seen in right panels of Figs.~\ref{fig:4} \& \ref{fig:5} and Table~\ref{tab:2}). The evolution of the single peak to double peak QPO was also seen in the dynamic PDS (Fig. \ref{fig:6}) with frequency difference $\delta f ~\sim12$ mHz between the two peaks. The static PDS is mostly dominated with flat-top noise at lower frequencies or the powerlaw nature in some of the observation segments. 

A few of the `heartbeat' QPOs have a $Q_\mathrm{factor}<3$ and amplitude unlike the typical values found in other BH sources \citep{Case2004}. These could be because the variability observed does not strictly conform to the $\rho$-class observed in previous outbursts. Given time, the light curve could evolve to regular, structured variability. However, due to lack of continuous observation of the source, further conclusions cannot be drawn. Besides these, we are also able to detect very low frequencies and harmonics in the range of 5--10 mHz and 1.25--2 mHz in some of the segments. Previous reports of such `heartbeat' QPOs exist, along with the detection of an QPO at 5 Hz \citep{Alta2011,Pahari2014}.

However, the \textit{LAXPC} PDS is too noise dominated beyond 1 Hz along with presence of some signature of GX 349+2 make it difficult to detect any such feature in contrast with GRS 1915+105 where a QPO at $\sim 5$ Hz is observed along with a sub-harmonic at $\sim 2.6^{+0.05}_{-0.06}$ Hz. Also, there is no conclusive evidence for double peaks in `heartbeat' QPO observed in the PDS of GRS 1915+105. \cite{Pahari2014} and \cite{Iyer2015} observed that the timescale of variability decreased as the source evolved into a `$\rho$'-class variability. This was also observed for the subsequent observations of GRS 1915+105 where timescale decreased from $\sim 150$ s to $\sim 100$ s \citep{Rawat2019}.
However, we observed that the variability timescales in IGR J17091--3624 seem to increase with time for segments where single peak QPOs are observed  (see Table \ref{tab:2}). In segments where double peak QPOs are observed, the two different timescales do not seem to follow any individual pattern. However, an overall increase in timescales is seen in these segments too. \cite{Chakra2000} state that the low frequency QPOs (1--10 mHz) are due to transitions between the burst and quiescent states. These low frequency QPOs observed in black hole binary systems provide the information of physical oscillations in the system. The observed change in the frequency might be associated with the change in the angular momentum of the incoming matter due to the interaction with the slowly inward propagating matter in the Keplerian disc or change of viscosity \citep{Das2014,Mond2015}, periodic orbital motion of companion or release of the matter at different conditions. It is also believed that a lower angular momentum accretion flow could be released from the magnetic trap inside the Roche lobe of the black hole binary systems \citep{Suko2017}.

Due to the possible contamination of \textit{LAXPC} data from the nearby source GX 349+2, broadband spectral analysis of IGR J17091--3624 was limited to the energy range of 0.7--23 keV. The spectra of both IGR J17091--3624 and GRS 1915+105 in the energy range 0.7--7 keV is dominated by a thermal disc with similar temperatures. However, the norm for both sources differs significantly while fitting with Model 1, with IGR J17091--3624 having a much smaller norm ($\sim 7$ compared to $\sim 40$ from GRS 1915+105). This implies a disc which could extend closer to the compact object. This is interesting as the estimated mass of the two sources lie in a similar range 10--14 $M_{\odot}$. 
The IGR J17091--3624 broadband spectrum analysis revealed that there is a minimal change in an average disc temperature with range of 0.63--0.84 keV with similar variation in electron plasma temperature having range of 3.5--4.0 keV. The photon index varying in the range of 2.13--2.40 are consistent with the source being in softer state. Slight variations in the spectral parameters is possible, since the exact spectral nature of GX 349+2 at the time of observation is unknown.

The centroid frequency of the `Heartbeat' QPO exhibits an overall decreasing trend. Decrease in rms from 29 to 17\% is seen with time (Table~\ref{tab:2}) accompanied by a corresponding decrease in $\Gamma$ from 2.4 to 2.1. Variation in the input seed photon temperature is also seen from 0.7--0.4 keV from segment 3 to segment 16. The time averaged spectral and temporal properties of IGR J17091-3624 are broadly consistent with the disc-truncation theory where the disc recedes from the compact object as it moves from SIMS to a harder state. However, significant differences in spectral properties in segments exhibiting SPQ and DPQ in PDS are absent (see Table~\ref{tab:4}). PRS was then performed to probe for further differences in these two segments. The broadband PRS of IGR J17091--3624 shows a clear difference in the disc temperature, electron plasma temperature and photon index in both phases I \& II as mention in the Table~\ref{tab:6} with respect to an average value mentioned in the Table~\ref{tab:4}. It was found that the parameters during the phase II in PRS were higher and in phase I well below the average value. Although differentiation of the spectral properties in the SPQ and DPQ segments was not possible due to marginal variation in the parameters. In GRS 1915+105, photon index ($\Gamma$) increases from $P_{a}$ to $P_{c}$  signaling a transition from harder to softer state. However, the increase in the disc temperature is not as high as seen in IGR J17091-3624. A significant variation in electron plasma temperature (13--32 keV) is also seen in the case of GRS 1915+105 compared to the near constant values observed in IGR J17091-3624. Although, accumulation of a larger data set for analysis is recommended before a physical interpretation can be attempted. The dominance of the disc component implies that the variability observed in both sources is basically thermal in nature during the `heartbeat' class. Further studies could present interesting constraints on the geometry and disc size of these sources.

The X-ray flux of IGR J17091--3624 was found to be $4.3 \times 10^{-9}$ ergs cm$^{-2}$ s$^{-1}$ in the energy range of 0.1--100 keV while that of GRS 1915+105 was obtained as $2.0 \times 10^{-8}$ ergs cm$^{-2}$ s$^{-1}$. The absorption column density is (0.8--1.1)$\times 10^{22}$ cm$^{-2}$ and $5.4 \times 10^{22}$ cm$^{-2}$ for IGR J17091--3624 and GRS 1915+105 respectively, which is similar to the those reported by \cite{Pahari2014}. Considering IGR J17091--3624 to be at an average distance of $14^{+3}_{-3}$ kpc, the luminosity is found to be $1.0\times 10^{38}$ erg s$^{-1}$, while for GRS 1915+105 at a distance of $8.5^{+2.0}_{-1.6}$ kpc, the luminosity is obtained as $1.7\times 10^{38}$ erg s$^{-1}$. IGR J17091--3624 seems to be emitting at 4--11\% of $L_\mathrm{Edd}$ while GRS 1915+105 emits at 6--20\% of $L_\mathrm{Edd}$ considering uncertainties in flux, distance and mass.
The $\rho$-class variabilities are generally attributed to the fast removal and refilling of matter in the inner accretion disc \citep{Belloni1997}, which is driven by radiation pressure instability in a standard Shakura-Sunyaev accretion disc \citep{Shakura1973} which requires the source to be emitting at close to $L_{Edd}$ to trigger the limit cycle instability \citep{Done2004}. However the luminosity obtained for both sources is much lower than the 70--80\% of $L_\mathrm{Edd}$ as obtained for $\rho$-class by \cite{Neilsen2011,Neilsen2012}. \cite{Rawat2019} suggest that the HS class in GRS 1915+105 could be a precursor to the $\rho$-class, which in some time may evolve to the $\rho$-class as defined by \cite{Bell2001}. Nevertheless, it is clear that `heartbeat' variability is observed in both sources although both of them emit at sub-Eddington luminosities. Earlier studies propose multiple reasons including spectral deformation due to higher inclination angle \citep{Pahari2014}, low or retrograde spin in IGR J17091--3624 \citep{Rao2012}, presence of a very small BH ($< 3M_{\odot}$) or alternatively the source being at a larger distance ($> 20$ kpc) \citep{Alta2011}, or a difference in mass accretion rate, for the detection of GRS 1915+105 like variabilities in IGR J17091--3624 which has always been observed at lower flux than for GRS 1915+105.  

We make use of the parameters obtained with the \texttt{ezdiskbb} model to estimate the mass accretion rate \citep{Frank2002} as,

\begin{equation}
	\dot{m}=\frac{8 \pi R_\mathrm{in,km}^{3} \sigma T_\mathrm{in,keV}^{4}}{3GM_{bh}} \;\;  M_{\odot} yr^{-1},
\end{equation}
where $\sigma$ is Boltzmann constant and $M_{bh}$ is the mass of the black hole and $R_\mathrm{in,km}$ is obtained as,
\begin{equation}
    R_\mathrm{in,km}=C^{2} \times \sqrt{N_\mathrm{ezdiskbb}} \times D_\mathrm{10,kpc}/ \sqrt{\cos i},
\end{equation}
where $N_\mathrm{ezdiskbb}$ is the norm of the \texttt{ezdiskbb} model, $D_\mathrm{10,kpc}$ is the distance to the source in units of 10 kpc, $i$ is the inclination angle and $C$ is the colour correction factor which we assume to be 1.7 for both sources \citep{Shim1995}. The disc radius is then obtained in the range 21--34 km and 13--24 km respectively for GRS 1915+105 and IGR J17091-3624 respectively.

We also obtain the ratio of mass accretion rate for IGR J17091--3624 and GRS 1915+105  for a distance range of 11--17 kpc and 6.9--10.5 kpc respectively. Considering the mass of IGR J17091--3624 and GRS 1915+105 as $11.5^{+0.8}_{-0.9}$ and $12.4^{+2.4}_{-1.8}$ $M_{\sun}$, we obtain a maximum $\dot{m}_\mathrm{GRS}/\dot{m}_\mathrm{IGR} \sim 25$ for an inclination angle of $70^{\circ}$ which is much lower than the value obtained by \cite{Pahari2014} at 48.3. This is expected as GRS 1915+105 is brighter than IGR J17091--3624 by only a factor close to 2. Also better estimates of mass and distance for these two sources are now available. Therefore, the condition that the source needs to emit at Eddington luminosities to exhibit `heartbeat' class variability does not seem like a sufficient condition given the low luminosities obtained for both sources ($>26\%$ of $L_\mathrm{Edd}$ which is required for the observed variabilities based on simulations by \cite{Nayakshin2000}). Based on this, \cite{Court2017} suggested that the size of the disc and the magnetic field required to stabilize such a disc in a radiation dominated field could contribute to the GRS 1915+105 like variabilities observed. Sub-Eddington accretion rates were observed for IGR J17091--3624 by \cite{Iyer2015}, where they suggest that the complex interplay between Keplerian and sub-Keplerian flows in the presence of outflows/winds could give rise to this variability \citep{Mandal2010,Radhi2018}.
\citet{Mass2014}, using a non-linear differential equations to reproduce the light curve and the energy lags in a $\rho$-state observed with BeppoSAX, suggested that the level and stability of the accretion rate could be the parameter responsible for the onset of the `heartbeat' oscillations. Further studies aiming at a consistent model which explains the origin of these variabilities need to be done to get a comprehensive picture.

\section*{Acknowledgments}

We thank the anonymous reviewer for the suggestions to improve the quality of the manuscript. We acknowledge all the support from the Indian Space Research Organization (ISRO) for \textit{AstroSat} mission, in particular, for providing the necessary software tools and for releasing the data through ISSDC data archive centre. We also acknowledge the support received from the \textit{LAXPC} and \textit{SXT} Payload Operation Centres (POC), TIFR, Mumbai for release of verified data, calibration data products and pipeline processing tools. 
We thank Radhika and co-authors for providing us with their analysis results, which was used to comment on the spectral state of the source.
BEB, VKA and AN thank GH, SAG; DD, PDMSA and Director, URSC for encouragement and continuous support to carry out this research. 
TK, HMA and KM acknowledge support of the Department of Atomic Energy, Government of India,
under project no.~12~R\&D-TFR-5.02-0200.

\section*{DATA AVAILABILITY STATEMENT}
Data underlying this article are available at AstroSat-ISSDC website (\url{http://astrobrowse.issdc.gov.in/astro_archive/archive}), MAXI website (\url{http://maxi.riken.jp/mxondem}) and Swift/BAT website (\url{https://swift.gsfc.nasa.gov/results/transients/}).


\bibliographystyle{mnras}
\bibliography{ref_grs} 
\bsp	
\label{lastpage}
\end{document}